\documentclass[a4paper,11pt]{article}%
\usepackage{amssymb}
\usepackage{graphicx}
\usepackage{amsbsy}
\usepackage{amsmath}
\usepackage{cite}
\usepackage{slashed}
\usepackage{amsfonts}
\usepackage{hyperref}
\hypersetup{
colorlinks   = true,
linkcolor    = blue,
citecolor    = blue,
}

\usepackage{color}
\definecolor{Blu}{rgb}{0.,0.,1.}

\begin{document}
\thispagestyle{empty} \setcounter{page}{0} \begin{flushright}
October 2020\\
\end{flushright}

\vskip          4.1 true cm

\begin{center}
{\huge Variations on the SU(5) axion}\\[1.9cm]

\textsc{J\'er\'emie Quevillon}$^{1}$\textsc{\ and Christopher Smith}$^{2}%
$\vspace{0.5cm}\\[9pt]\smallskip{\small \textsl{\textit{Laboratoire de
Physique Subatomique et de Cosmologie, }}}\linebreak%
{\small \textsl{\textit{Universit\'{e} Grenoble-Alpes, CNRS/IN2P3, Grenoble
INP, 38000 Grenoble, France}.}} \\[1.9cm]\textbf{Abstract}\smallskip
\end{center}

\begin{quote}

The simultaneous embeddings of an axion state and a seesaw mechanism within the $SU(5)$ Grand Unification Theory, both minimal and flipped, are systematically studied. It is shown that whenever $\mathcal{B}-\mathcal{L}$ is active as a global symmetry, the PQ charges of the fermions are ambiguous. Various realizations of the seesaw mechanism are then found to differ only in the way this ambiguity is resolved. But since the ambiguity is unphysical, all these models are thereby predicted to be equivalent, in the sense that the axion phenomenology is identical at the low energy scale.

\let\thefootnote\relax\footnotetext{$^{1}$ jeremie.quevillon@lpsc.in2p3.fr}\footnotetext{$^{2}$ chsmith@lpsc.in2p3.fr}
\end{quote}

\newpage

\setcounter{tocdepth}{3}
\tableofcontents

\section{Introduction}

Axions are probably the best solutions for the strong CP problem of the Standard Model (SM). This problem originates from the observation that the electroweak and the QCD sectors, by construction
secluded, must conspire to cancel each other's contribution to the electric dipole moment of the neutron to an impressive precision of about one part in $10^{10}$~\cite{Abel:2020gbr}.

The fundamental building block of all axion models is first a spontaneously broken $U(1)_{PQ}$ symmetry, the Peccei-Quinn (PQ) symmetry, and second some colored chiral fermions charged under that symmetry~\cite{PQ}. This makes the symmetry anomalous, and ensures the Goldstone boson~\cite{Weinberg:1977ma,Wilczek:1977pj} arising from the $U(1)_{PQ}$ breaking, the axion, is anomalously coupled to gluons. Subsequently, out of this gluonic coupling, non-perturbative QCD effects create an effective potential for the axion field, such that the strong CP puzzle disappears precisely when the axion falls at the minimum. In the process, the axion acquires a small mass, typically well below the eV scale~\cite{Bardeen:1977bd,Kim:1986ax}. Both its mass and its couplings are thus controlled by the single scale $f_{a}$ of the spontaneous symmetry breaking of $U(1)_{PQ}$. Constraints from astrophysics and particle physics call for this scale to be much larger than the electroweak scale, $f_a > 10^{9}$~GeV~\cite{Dicus:1979ch}. 

Naturally it is tempting to identify the PQ breaking scale with the scale of grand unification, using the same field to break both symmetries, either in non-supersymmetric or supersymmetric contexts. The first attempt at embedding the axion in a GUT context was proposed by Dine, Fisher, and Srednicki in Ref.~\cite{axionDFS} and by Wise, Georgi and Glashow in Ref.~\cite{Wise:1981ry}, both in the context of the minimal non-supersymmetric $SU(5)$ model of Ref.~\cite{Georgi:1974sy}. Since then, additional studies of the embedding of the axion in non-supersymmetric grand unified theories have been proposed, see e.g. Refs.~\cite{FileviezPerez:2019fku,FileviezPerez:2019ssf} for $SU(5)$, or Refs.~\cite{Reiss:1981nd,Mohapatra:1982tc,Holman:1982tb,Bajc:2005zf,Bertolini:2012im,Altarelli:2013aqa,Babu:2015bna,Ernst:2018bib,Ernst:2018rod,DiLuzio:2020xgc} for $SO(10)$.

The PQ scale also looks suspiciously close to the seesaw scale, that is relevant to explain the small neutrino masses. So, it appears desirable to describe the three mechanisms -gauge coupling unification, small neutrino mass, axion solution to the strong CP puzzle- in a unified setting. The purpose of the present paper is to study such constructions, in the non-supersymmetric context. There have been a few attempts along this line, most notably Ref.~\cite{Bajc:2006ia} where a Majorana fermion in the adjoint representation of $SU(5)$ is used (see also Refs.~\cite{Bajc:2007zf,DiLuzio:2013dda,DiLuzio:2018gqe,Dorsner:2005fq}). Our purpose here is to study as systematically as possible the models based on $SU(5)$ and including both an axion and a seesaw mechanism. 

In the present analysis, we will extensively rely on the results of our recent studies in Refs.~\cite{Quevillon:2019zrd, Quevillon:2020hmx}. Indeed, there, we have shown in a non GUT context that whenever the SM fermions carry non trivial PQ charges, the PQ symmetry becomes entangled with the accidental $U(1)$ symmetries of the SM, corresponding to the conserved baryon ($\mathcal{B}$) and lepton ($\mathcal{L}$) numbers. This leads to ambiguities in the PQ charges of the fermions. Those have no phenomenological consequences, but become crucial to accommodate for additional $\Delta\mathcal{B}$ and/or $\Delta\mathcal{L}$ effects, and in particular, to allow for a seesaw mechanism. Further, since accounting for these effects simply fixes some ambiguous thus unphysical parameters, it is immediately clear that the axion phenomenology is totally unaffected. In this way, seemingly different axion models can be shown to be impossible to distinguish phenomenologically.

In a $SU(5)$ context, the entanglement of $U(1)_{PQ}$ with $U(1)_{\mathcal{B},\mathcal{L}}$ must take a particular form since $\mathcal{B}+\mathcal{L}$ does not survive as a global symmetry. So, our first goal will be to precisely identify the PQ charge ambiguity present in models where only $\mathcal{B}-\mathcal{L}$ is active. Then, we will check whether this ambiguity permits to accommodate for $\mathcal{L}$-breaking seesaw mechanisms of various types. Further, we will follow this strategy both for the minimal $SU(5)$ and for the flipped $SU(5)$ GUT model~\cite{Barr:1981qv,Derendinger:1983aj,Antoniadis:1987dx, Ellis:1988tx}, and consider various alternative embedding of the axion in $SU(5)$ representations. All these seemingly very different models, in which the fermions do have very different PQ charges, will be shown to be mere particular cases of the $\mathcal{B}-\mathcal{L}$ preserving models, with the ambiguity fixed to some particular value. As a result, and despite their very different appearance in terms of effective interactions, those models cannot be distinguished at low energy.

The paper is organized as follow. To set the stage, we start in the next section by briefly summarizing a few relevant features the PQ axion model and the DFSZ axion model (more details can be found in Ref.~\cite{Quevillon:2019zrd, Quevillon:2020hmx}). Then in Section~\ref{sec3}, we study the minimal $SU(5)$ axion models compatible with lepton number violation, by introducing various mechanisms to generate neutrino masses but also baryon number violation. In Section~\ref{sec4} we investigate the Flipped $SU(5)$ axion models, characterize the particular way in which $\mathcal{B}-\mathcal{L}$ is broken (or not), and two different DFSZ implementations for the axion. Finally, our results are summarized in Section~\ref{Ccl}. A few remarks about how these models necessarily have to be modified to reproduce realistic fermion masses are collected in the Appendix.

\section{Brief overview of the PQ and DFSZ axions}

In the presence of two Higgs doublets $\Phi_{1,2}$, the whole Lagrangian can be required to be invariant under a global $U(1)_{1}\otimes U(1)_{2}$ symmetry, corresponding to the independent rephasing of each doublet, $\Phi_{k}\rightarrow \exp (i \alpha_k)\Phi_{k}$. This imposes some restrictions on the scalar potential and on the Yukawa couplings, which we take to be of Type II,%
\begin{equation}
\mathcal{L}_{\text{Yukawa}}=-\bar{u}_{R}\mathbf{Y}_{u}q_{L}\Phi_{1}-\bar
{d}_{R}\mathbf{Y}_{d}q_{L}\Phi_{2}^{\dagger}-\bar{e}_{R}\mathbf{Y}_{e}\ell
_{L}\Phi_{2}^{\dagger}+h.c.\;.\label{YukQuark}%
\end{equation}
Because these couplings are also invariant under the global baryon and lepton number symmetries, $U(1)_{\mathcal{B}}$ and $U(1)_{\mathcal{L}}$, the pattern of symmetry breaking is%
\begin{align}
G_{THDM}  & =U(1)_{\mathcal{B}}\otimes U(1)_{\mathcal{L}}\otimes
U(1)_{1}\otimes U(1)_{2}\otimes SU(2)_{L}\otimes SU(3)_{C}\nonumber\\
& \rightarrow U(1)_{\mathcal{B}}\otimes U(1)_{\mathcal{L}}\otimes
U(1)_{em}\otimes SU(3)_{C} \ ,
\label{PatternTHDM}%
\end{align}
where $U(1)_Y\subset U(1)_{1}\otimes U(1)_{2}$ is gauged. When the doublets acquire their vacuum expectation values, $\langle 0|\operatorname{Re}\Phi_{i}|0\rangle=v_{i}$ with $v_{1}^{2}+v_{2}^{2}\equiv v^{2}\approx\left(  246\,\text{GeV}\right)  ^{2}$ and $v_{2}/v_{1}\equiv x\equiv1/\tan\beta$, the symmetry $U(1)_{1}\otimes U(1)_{2}\otimes SU(2)_{L}$ is broken down to $U(1)_{em}$. There are two electrically-neutral Goldstone bosons~\cite{Weinberg:1977ma,Wilczek:1977pj}: the Would-be Goldstone eaten by the $Z^{0}$ and the massless axion. Importantly, these states are only defined once $U(1)_{Y}\otimes SU(2)_{L}$ is broken, and are $v_{i}$-dependent linear combinations of $\operatorname{Im}\Phi_{1}$ and $\operatorname{Im}\Phi_{2}$. The PQ charges of the Higgs doublets are also function of the VEVs, and the orthogonality of the Goldstone bosons imposes%
\begin{equation}
PQ(\Phi_{1},\Phi_{2})=\left(  x\ ,\ -\frac{1}{x}\ \right)  \ .
\end{equation}
These charges fix those of the fermions, up to a two-parameter ambiguity originating in the $U(1)_{\mathcal{B}}\otimes U(1)_{\mathcal{L}}$ invariance, which we denote $\alpha$ and $\beta$:%
\begin{equation}
PQ(q_{L},u_{R},d_{R},\ell_{L},e_{R})=(\alpha,\alpha+x,\alpha+\frac{1}{x}%
,\beta,\beta+\frac{1}{x})\ \ .\label{PQferm}%
\end{equation}
As detailed in Refs.~\cite{Quevillon:2019zrd,Quevillon:2020hmx}, the freedom in the PQ charge of the fermion has no observable consequence. Yet, some theoretical quantities depend on $\alpha$ and $\beta$. In particular, the divergence of the PQ current takes the form
\begin{equation}
\partial_{\mu}J_{PQ}^{\mu}=\frac{N_{f}}{16\pi^{2}}\left\{  \mathcal{N}%
_{C}g_{s}^{2}G_{\mu\nu}^{a}\tilde{G}^{a,\mu\nu}+\mathcal{N}_{L}g^{2}W_{\mu\nu
}^{i}\tilde{W}^{i,\mu\nu}+\mathcal{N}_{Y}g^{\prime2}B_{\mu\nu}\tilde{B}%
^{\mu\nu}\right\}  \;,\label{dJPQTHDM}%
\end{equation}
with
\begin{subequations}
\label{JPQTHDM}%
\begin{align}
\mathcal{N}_{C} &  =\sum_{\ \ \ \psi=q_{L}^{\dagger},u_{R},d_{R}\ \ \ }%
d_{L}(\psi)C_{C}(\psi)PQ(\psi)=\frac{1}{2}\left(  x+\frac{1}{x}\right)  \ ,\\
\mathcal{N}_{L} &  =\sum_{\ \ \ \ \ \psi=q_{L}^{\dagger},\ell_{L}^{\dagger
}\ \ \ \ \ }d_{C}(\psi)C_{L}(\psi)PQ(\psi)=-\frac{1}{2}(3\alpha+\beta)\ ,\\
\mathcal{N}_{Y} &  =\sum_{\psi=q_{L}^{\dagger},u_{R},d_{R},\ell_{L}^{\dagger
},e_{R}}d_{L}(\psi)d_{C}(\psi)C_{Y}(\psi)PQ(\psi)=\frac{1}{2}\left(
3\alpha+\beta\right)  +\frac{4}{3}\left(  x+\frac{1}{x}\right)  \ ,
\end{align}
where $d_{L(C)}$, $c_{L(C)}$ are the $SU(2)_{L}$ ($SU(3)_{C}$) dimension and quadratic Casimir invariant, and $C_{Y}=Y^{2}/4$. With $\mathcal{N}_{L}+\mathcal{N}_{Y}=\mathcal{N}_{em}$, the QED and QCD terms~\cite{PQ} in $\partial_{\mu}J_{PQ}^{\mu}$ are physical but the electroweak term is ambiguous due to the presence of the free parameters, $\alpha$ and $\beta$.

With the axion emerging from the Higgs doublets, its couplings to SM particles are tuned by the electroweak VEV, and are phenomenologically too large. To circumvent this, the idea of the DFSZ model~\cite{axionDFS,axionZ} is to embed the axion dominantly in a separate complex scalar field $\phi$, whose VEV $v_{s}$ is much larger than the electroweak
one. Technically, the introduction of the complex scalar field does not enlarge the $U(1)_{1}\otimes U(1)_{2}$ symmetry thanks to the presence of a coupling $\phi^{2}\Phi_{1}^{\dagger}\Phi_{2}$ entangling the charges of all the scalars. This also prevents $\phi$ from coupling to fermions. The axion emerges as essentially $\operatorname{Im}\phi$, with small $\mathcal{O}(v/v_{s})$ components $\operatorname{Im}\Phi_{1,2}$. Since all the couplings to SM particles stem from these suppressed components, the axion couplings are all rescaled by $v/v_{s}$.

The PQ charges of the doublets are not modified by the presence of $\phi$. Since it has no weak hypercharge, it does not enter in the WBG of the $Z^{0}$ to which the axion must be orthogonal, and thus:
\end{subequations}
\begin{equation}
PQ(\Phi_{1},\Phi_{2},\phi)=\left(  x\ ,\ -\frac{1}{x}\ ,\ \frac{1}{2}\left(
x+\frac{1}{x}\right)  \right)  \ .\label{PQdfsz}%
\end{equation}
Also, the SM fermion PQ charges remain those of Eq.~(\ref{PQferm}) since the
Yukawa couplings are the same.

\section{Minimal SU(5) axion models\label{sec3}}

Since phenomenological constraints push the invisible axion scale well above the electroweak scale, it could be related to other new physics scales, in particular to the seesaw scale of the neutrino sector or the grand unification scale suggested by the RG evolution of the SM gauge couplings. With these contexts in mind, the goal would actually be to relate these three scales, and this section is devoted to such constructions.

At first sight, the PQ and DFSZ axions look incompatible with unification because the PQ charges for the fermions embedded in the same $SU(5)$ representation are different. Yet, as we will explain here, this line of reasoning is flawed, and the very idea of the DFSZ axion can be transposed quite naturally in $SU(5)$. One may also consider a KSVZ-like axion~\cite{KSVZ}, but having to embed the heavy fermion into a complete
$SU(5)$ representation requires extending the matter content quite
extensively. This will not be explored here.

In the next section, the minimal $SU(5)$ model is briefly presented, focusing
on those elements that will play a role in the following. Then, we extend this
model in various ways to include the axion. In that description, particular
emphasis is laid on global symmetries and the breaking chains. Indeed, in
a GUT setting, the $\mathcal{B}$ and $\mathcal{L}$ are not exact
symmetries at the GUT scale, but only emerge at the low scale. As we will see, this means the PQ symmetry should be defined similarly if it is to be compatible with $\mathcal{B}$ and/or $\mathcal{L}$ violating effects, as required for example to allow for a Majorana neutrino mass term.

\subsection{Brief overview of the minimal SU(5) model}

The simplest GUT model, due to Georgi and Glashow \cite{Georgi:1974sy}, is based on $SU(5)$. Fermions are embedded in the fundamental representations $\psi_{\mathbf{\bar{5}}}\sim\mathbf{\bar{5}}$ and $\chi_{\mathbf{10}}\sim\mathbf{10}$, while gauge bosons are in the adjoint, $A^{\mu}\sim\mathbf{24}$. Two Higgs multiplets are necessary to break $SU(5)$ down to $SU(3)_{C}\otimes U(1)_{em}$: a set of real scalar fields $\mathbf{H}_{\mathbf{24}}\sim\mathbf{24}$ responsible for $SU(5)\rightarrow SU(3)_{C}\otimes SU(2)_{L}\otimes U(1)_{Y}$ at the GUT scale $v_{24}$, and a complex fiveplet $h_{\mathbf{5}}\sim\mathbf{5}$ for the EW symmetry breaking at the scale $v_{5}\approx 246\,$GeV. The most general scalar potential, assuming a $\mathbf{H}_{\mathbf{24}}\rightarrow-\mathbf{H}_{\mathbf{24}}$ symmetry to get rid of cubic interactions, is%
\begin{align}
V(h_{\mathbf{5}},\mathbf{H}_{\mathbf{24}}) &  =-\frac{\mu^{2}}{2}%
\langle\mathbf{H}_{\mathbf{24}}^{2}\rangle+\frac{a}{4}\langle\mathbf{H}%
_{\mathbf{24}}^{2}\rangle^{2}+\frac{b}{2}\langle\mathbf{H}_{\mathbf{24}}%
^{4}\rangle\nonumber\\
&  -\frac{\mu^{\prime2}}{2}h_{\mathbf{5}}^{\dagger}h_{\mathbf{5}}%
+\frac{\lambda}{4}(h_{\mathbf{5}}^{\dagger}h_{\mathbf{5}})^{2}+\alpha
(h_{\mathbf{5}}^{\dagger}h_{\mathbf{5}})\langle\mathbf{H}_{\mathbf{24}}%
^{2}\rangle+\beta h_{\mathbf{5}}^{\dagger}\mathbf{H}_{\mathbf{24}}%
^{2}h_{\mathbf{5}}\;,
\end{align}
and it can achieve the desired symmetry breaking chain for some appropriate choices of the parameters.

Two features of the minimal model need to be emphasized. First, the fermion masses are not correctly predicted by the minimal model. From the Yukawa couplings%
\begin{equation}
\mathcal{L}_{\text{Yukawa}}^{\mathbf{5}}=-\frac{1}{4}\varepsilon_{ABCDE}(\bar{\chi
}_{\mathbf{10}}^{c})^{AB}\mathbf{Y}_{10}(\chi_{\mathbf{10}}%
)^{CD}h_{\mathbf{5}}^{E}+\sqrt{2}(\bar{\psi}_{\mathbf{\bar{5}}}^{c%
})_{A}\mathbf{Y}_{5}(\chi_{\mathbf{10}})^{AB}(h_{\mathbf{5}}^{\dagger}%
)_{B}+h.c.\;,
\end{equation}
one derives%
\begin{equation}
\mathbf{Y}_{u}=\mathbf{Y}_{10}=\mathbf{Y}_{10}^{T}\;,\;\;\mathbf{Y}%
_{d}=\mathbf{Y}_{e}^{T}=\mathbf{Y}_{5}\;.\label{GUTminmass}%
\end{equation}
To cure for this, either effective non-renormalizable operators are added, or a non-minimal Higgs multiplet is included. These constructions are briefly described in the Appendix~\ref{AppMass}. 

A second important feature of the minimal $SU(5)$ model is the existence of a global $U(1)_{X}$ symmetry, with charges
\begin{equation}
X(h_{\mathbf{5}})=-2X(\chi_{\mathbf{10}})=X(\chi
_{\mathbf{10}})+X(\psi_{\mathbf{\bar{5}}})\;.
\end{equation}
The $\mathbf{H}_{\mathbf{24}}$ and gauge bosons are neutral. This symmetry has an interesting property. It is a subgroup of the $U(1)_{5}\otimes U(1)_{10}$ symmetry corresponding to the separate rephasing of $\psi_{\mathbf{\bar{5}}}$ and $\chi_{\mathbf{10}}$. Both the $U(1)_{5}$ and $U(1)_{10}$ singlet currents are chiral and thus anomalous:
\begin{equation}
\left(
\begin{array}
[c]{c}%
\partial_{\mu}J_{\bar{5}}^{\mu}\\
\partial_{\mu}J_{10}^{\mu}%
\end{array}
\right)  =-N_{f}\frac{g_{5}^{2}}{16\pi^{2}}\left(
\begin{array}
[c]{c}%
1/2\\
3/2
\end{array}
\right)  A_{\mu\nu}^{A}\tilde{A}^{A,\mu\nu}\;,\label{AnoSU5}%
\end{equation}
where $C(\mathbf{\bar{5}})=1/2$, $C(\mathbf{10})=3/2$, $N_{f}=3$ the number of fermion generations, and $\tilde{A}^{A,\mu\nu}=1/2\varepsilon^{\mu\nu\rho\sigma}A_{\rho\sigma}^{A}$. However, the combination $J_{X}^{\mu}\equiv J_{10}^{\mu}-3J_{\bar{5}}^{\mu}$ corresponding to the fermionic current of the $U(1)_{X}$ symmetry is anomaly-free.

The $U(1)_{X}$ symmetry is not broken by the VEV of $\mathbf{H}_{\mathbf{24}}$ since that state is neutral. It is only broken at the electroweak scale, along with $U(1)_{Y}$, once $h_{\mathbf{5}}$ develops its VEV. There is no associated Goldstone boson because this actually corresponds to a partial breaking, with a reordering of the $U(1)$s:%
\begin{equation}
SU(2)_{L}\otimes U(1)_{Y}\otimes U(1)_{X}\rightarrow U(1)_{em}\otimes
U(1)_{\mathcal{B}-\mathcal{L}}\ .
\end{equation}
In other words, there is no Goldstone boson associated with the $U(1)_{X}$ breaking because it gets replaced by another exact global symmetry~\cite{Mohapatra:1982xz}. To see that the surviving $U(1)$ is actually that for $\mathcal{B}-\mathcal{L}$, first note that the $h_{\mathbf{5}}$ breaks both $U(1)_{Y}$ and $U(1)_{X}$,
but not the combination%
\begin{equation}
Z=\frac{1}{5}X+\frac{2}{5}Y\;,
\end{equation}
if we normalize the $X$ charge as $X(h_{\mathbf{5}})=-2$. It is then a simple exercise to check that $Z$ charges coincide with $\mathcal{B}-\mathcal{L}$ for the fermionic $SU(5)$ fields. Some of the GUT scale bosons also carry $\mathcal{B}-\mathcal{L}$ charges: the leptoquarks $X^{\mu}$ and $Y^{\mu}$ have $\mathcal{B}-\mathcal{L}=2/3$ and the colored states in $h_{\mathbf{5}}$ have $\mathcal{B}-\mathcal{L}=-2/3$.

\subsection{The PQ-SU(5) model}

There is no room for the axion in the minimal model, where no new matter fields are introduced. The simplest way to introduce it is to apply the PQ recipe~\cite{PQ} and add a second Higgs fiveplet. The Yukawa couplings are then%
\begin{equation}
\mathcal{L}_{\text{Yukawa}}=\frac{1}{4}\bar{\chi}_{\mathbf{10}}^{c%
}\mathbf{Y}_{10}\chi_{\mathbf{10}}h_{1,\mathbf{5}}+\sqrt{2}\bar{\psi
}_{\mathbf{\bar{5}}}^{c}\mathbf{Y}_{5}\chi_{\mathbf{10}%
}h_{2,\mathbf{5}}^{\dagger}+h.c.\;.\label{SU5THDM}%
\end{equation}
We want the Lagrangian to be invariant under $U(1)_{1}\otimes U(1)_{2}$ corresponding to the independent phase redefinitions of the two Higgs fiveplets $h_{k,\mathbf{5}}\rightarrow\exp(i\alpha_{k})h_{k,\mathbf{5}}$, so
the potential is restricted to%
\begin{align}
V(h_{1,\mathbf{5}},h_{2,\mathbf{5}},\mathbf{H}_{\mathbf{24}}) &  =-\frac
{\mu^{2}}{2}\langle\mathbf{H}_{\mathbf{24}}^{2}\rangle+\frac{a}{4}%
\langle\mathbf{H}_{\mathbf{24}}^{2}\rangle^{2}+\frac{b}{2}\langle
\mathbf{H}_{\mathbf{24}}^{4}\rangle\nonumber\\
&  \ \ \ -\sum_{i=1,2}\frac{\mu_{i}^{2}}{2}h_{i,\mathbf{5}}^{\dagger
}h_{i,\mathbf{5}}+\alpha_{i}(h_{i,\mathbf{5}}^{\dagger}h_{i,\mathbf{5}%
})\langle\mathbf{H}_{\mathbf{24}}^{2}\rangle+\beta_{i}h_{i,\mathbf{5}%
}^{\dagger}\mathbf{H}_{\mathbf{24}}^{2}h_{i,\mathbf{5}}+\frac{\lambda_{i}}%
{2}(h_{i,\mathbf{5}}^{\dagger}h_{i,\mathbf{5}})^{2}\nonumber\\
&  \ \ \ +\lambda_{3}(h_{1,\mathbf{5}}^{\dagger}h_{1,\mathbf{5}}%
)(h_{2,\mathbf{5}}^{\dagger}h_{2,\mathbf{5}})+\lambda_{4}(h_{1,\mathbf{5}%
}^{\dagger}h_{2,\mathbf{5}})(h_{2,\mathbf{5}}^{\dagger}h_{1,\mathbf{5}})\ .
\end{align}

The symmetry breaking proceeds similarly as in the minimal model. The parameters $\mu$, $a$, and $b$ can be chosen such that $\langle0|\mathbf{H}_{\mathbf{24}}|0\rangle=v_{24}\,\operatorname{diag}(1,1,1,-3/2,-3/2)$. Then, provided $\beta_{i}$ are negative, the terms $\beta_{i}h_{i,\mathbf{5}}^{\dagger}\langle0| \mathbf{H}_{\mathbf{24}}|0\rangle^{2}h_{i,\mathbf{5}}$ tilt the potential so that minima of the form $\langle0|h_{i,\mathbf{5}}|0\rangle\sim(0,0,0,v_{51},v_{52})$ are lower than color-breaking ones. The $\beta_{i}$ terms also correct the initial $SU(5)$ breaking, with generically $\langle0|\mathbf{H}_{\mathbf{24}}|0\rangle=v_{24}\, \operatorname*{diag}(1,1,1,-(3+\epsilon)/2,-(3-\epsilon)/2)$, but this has a negligible impact at the EW scale. Neglecting all $\mathcal{O}(v_{5}/v_{24})$ corrections, the EW symmetry breaking induced by $\langle0|h_{i,\mathbf{5}}|0\rangle$ proceeds exactly as in the THDM. Broadly speaking, one can see that $\mu_{i}^{2}$, $\alpha_{i}$, and $\beta_{i}$ combine into a quadratic term for $h_{i,\mathbf{5}}$, while the $\lambda$ parameters simply match onto those of the THDM.

The axion arises from the phases of $h_{i,\mathbf{5}}$. If we denote $\langle0|h_{i,\mathbf{5}}|0\rangle\sim(0,0,0,0,v_{i})^{T}$ with $v_{1}^{2}+v_{2}^{2}=v_{5}^{2}$ and $x=v_{2}/v_{1}$, the identification of the $U(1)_{Y}$ Goldstone boson and of the axion proceeds exactly as in the THDM, and the fiveplets have the PQ charges:%
\begin{equation}
PQ(h_{1,\mathbf{5}})=x\ \ ,\ \ PQ(h_{2,\mathbf{5}})=-\frac{1}{x}%
\ .\label{PQSU5naif1}%
\end{equation}
If we interpret those charges at the level of the $SU(5)$ invariant Yukawa couplings of Eq.~(\ref{SU5THDM}), we find%
\begin{equation}
PQ(\chi_{\mathbf{10}})=-\frac{x}{2}\ ,\ \ PQ(\psi_{\mathbf{\bar{5}}})=\frac
{x}{2}-\frac{1}{x}\ .\label{PQSU5naif2}%
\end{equation}
The charge of $\chi_{\mathbf{10}}$ is fixed because of the Majorana nature of the $\mathbf{Y}_{10}$ coupling, and this then leaves no freedom for $\psi_{\mathbf{\bar{5}}}$. Note that if we start from the PQ charges of the fermions in the THDM, Eq.~(\ref{PQferm}), and require the equality of the charges of the fermions of the
$\mathbf{10}$ and $\mathbf{\bar{5}}$, we recover the above result with
\begin{equation}
\alpha=-\frac{x}{2}\ ,\ \ \beta=\frac{x}{2}-\frac{1}{x}\ .\label{UnifPQ}%
\end{equation}

It may seem the $SU(5)$ symmetry unambiguously fixes all the PQ charges, but this is suspicious. First, these charges are function of $x=v_{2}/v_{1}$, which is clearly defined only once the fiveplets acquire their VEVs, that is, at the very end of the breaking chain. Further, since the axion state knows about $U(1)_{Y}$ and its breaking, and since the hypercharge is not constant over $SU(5)$ multiplets, it is far from clear that the PQ charge should be constant over whole $SU(5)$ multiplets.

To get a better understanding, let us analyze the breaking chain in more details. In the THDM, the Lagrangian is invariant under $U(1)_{1}\otimes U(1)_{2}\otimes U(1)_{\mathcal{B}}\otimes U(1)_{\mathcal{L}}$, with $U(1)_{Y}$ hidden inside $U(1)_{1}\otimes U(1)_{2}$. With the breaking chain of Eq.~(\ref{PatternTHDM}), there are two Goldstone bosons, one is the axion corresponding to $U(1)_{PQ}$ and the other is eaten by the $Z$ boson. The third $U(1)$ remains exact, $U(1)_{em}\subset U(1)_{Y}\otimes SU(2)_{L}$. Thus, in the THDM, there is a two-parameter freedom in the PQ charges of the fermions because $U(1)_{\mathcal{B}}\otimes U(1)_{\mathcal{L}}$ is always exact and separate. The situation is quite different in $SU(5)$. Adding the
second Higgs fiveplet extends $U(1)_{X}$ to $U(1)_{1}\otimes U(1)_{2}$, and the breaking chain is%
\begin{align}
G_{SU(5)} &  \sim U(1)_{1}\otimes U(1)_{2}\otimes SU(5)\nonumber\\
&  \rightarrow U(1)_{1}\otimes U(1)_{2}\otimes U(1)_{Y}\otimes SU(2)_{L}%
\otimes SU(3)_{C}\nonumber\\
&  \sim U(1)_{X^{\prime}}\otimes U(1)_{X}\otimes U(1)_{Y}\otimes
SU(2)_{L}\otimes SU(3)_{C}\nonumber\\
&  \rightarrow U(1)_{\mathcal{B}-\mathcal{L}}\otimes U(1)_{em}\otimes
SU(3)_{C}\ .\label{SSB5}%
\end{align}
In the second stage of breaking, the $U(1)_{X}\subset U(1)_{1}\otimes U(1)_{2}$ mixes with $U(1)_{Y}\subset SU(5)$ to generate the unbroken $U(1)_{\mathcal{B}-\mathcal{L}}$, while $U(1)_{X^{\prime}}=U(1)_{1}\otimes
U(1)_{2}\backslash U(1)_{X}$ mixes with $U(1)_{Y}$ to generate the broken PQ symmetry. Thus, only one $U(1)\subset U(1)_{1}\otimes U(1)_{2}$ is spontaneously broken and only one additional Goldstone boson emerges. At the same time, though $U(1)_{\mathcal{B}-\mathcal{L}}$ is not explicitly present above the GUT scale, there is always a separate global symmetry that is active, and thus, we do expect the PQ charge of the fermions to exhibit a one-parameter freedom. This contradicts the naive Eq.~(\ref{UnifPQ}) where PQ charged are unambiguously fixed.

To pinpoint this freedom, let us start at the $SU(5)$ level. What matters there is the $U(1)_{1}\otimes U(1)_{2}$ symmetry, whose charges are%
\begin{equation}%
\begin{tabular}
[c]{ccccc}\hline
& $h_{1,\mathbf{5}}$ & $h_{2,\mathbf{5}}$ & $\chi_{\mathbf{10}}$ &
$\psi_{\mathbf{\bar{5}}}$\\\hline
$\ \ U(1)_{1}\ \ $ & $1$ & $0$ & $-1/2$ & $1/2$\\
$\ \ U(1)_{2}\ \ $ & $0$ & $1$ & $0$ & $1$\\\hline
\end{tabular}
\ \label{U1U2charges}%
\end{equation}
On the other hand, both PQ and $\mathcal{B}-\mathcal{L}$ charges are only
defined once $SU(5)$ is broken because of the mixing with $U(1)_{Y}$. Let us
impose the ansatz:%
\begin{align}
PQ &  =\frac{2}{5}(\zeta_{1}^{PQ}U_{1}+\zeta_{2}^{PQ}U_{2}+\zeta_{Y}%
^{PQ}Y)\ ,\label{PQU1U2Y}\\
\mathcal{B}-\mathcal{L} &  =\frac{2}{5}(\zeta_{1}^{\mathcal{B}-\mathcal{L}%
}U_{1}+\zeta_{2}^{\mathcal{B}-\mathcal{L}}U_{2}+\zeta_{Y}^{\mathcal{B}%
-\mathcal{L}}Y)\ ,\label{BLU1U2Y}%
\end{align}
for some coefficients $\zeta^{PQ}_{i}$ and $\zeta_{i}^{\mathcal{B}-\mathcal{L}}$. The factors $2/5$ are introduced for convenience. Remember that implicitly, PQ charges assume that $U(1)_{Y}$ is spontaneously broken since the PQ Goldstone boson is defined as that orthogonal to the WBG of the $Z^{0}$ boson. This fixes the PQ charges of the $SU(2)_{L}$ doublets as (compare with
Eq.~(\ref{PQSU5naif1}))%
\begin{equation}
PQ(\Phi_{1}\subset h_{1,\mathbf{5}})=x\ ,\ \ PQ(\Phi_{2}\subset
h_{2,\mathbf{5}})=-\frac{1}{x} \ ,
\end{equation}
but the color triplets in $h_{i,\mathbf{5}}$ may have different PQ charges. Then, imposing
Eq.~(\ref{PQferm}) and solving for $\alpha$, $\beta$, and $\zeta_{1,2,Y}^{PQ}$, a one-parameter freedom remains:
\begin{equation}
\zeta_{1}^{PQ}=\beta+2x+\frac{1}{x}\ ,\ \ \zeta_{2}^{PQ}=\beta-\frac{x}%
{2}-\frac{3}{2x}\ ,\ \ \zeta_{Y}^{PQ}=\frac{x}{2}-\frac{1}{x}-\beta
\ ,\label{PQU1U2Y2}%
\end{equation}
with $\beta$ arbitrary, together with%
\begin{equation}
3\alpha+\beta=-\left(  x+\frac{1}{x}\right)  \equiv2\mathcal{N}_{SU(5)}%
\ .\label{NSU5fixed}%
\end{equation}
So, we represent the PQ charge of the fermion keeping track of the
one-parameter freedom as%
\begin{equation}
PQ(q_{L},u_{R},d_{R},\ell_{L},e_{R})=\left(  \frac{2\mathcal{N}_{SU(5)}-\beta
}{3},\frac{2\mathcal{N}_{SU(5)}-\beta}{3}+x,\frac{2\mathcal{N}_{SU(5)}-\beta
}{3}+\frac{1}{x},\beta,\beta+\frac{1}{x}\right)  \ .\label{PQFinalCharge}%
\end{equation}
Written in this form, it is clear that the freedom $\beta$ corresponds to $\mathcal{B}-\mathcal{L}$ remaining exact. So, and quite expectedly, the only difference compared to the THDM is that the freedom corresponding to
$\mathcal{B}+\mathcal{L}$ no longer appears. Actually, it is the specific way in which the SU(5) gauge interactions break $\mathcal{B}+\mathcal{L}$ that freezes the combination $3\alpha+\beta$ to the specific value $2\mathcal{N}_{SU(5)}$.

Four features of this solution are remarkable. First, solving Eq.~(\ref{BLU1U2Y}) produces
\begin{equation}
\zeta_{Y}^{\mathcal{B}-\mathcal{L}}=-\zeta_{1}^{\mathcal{B}-\mathcal{L}%
}=-\zeta_{2}^{\mathcal{B}-\mathcal{L}}=1\ .\label{BLU1U2Y2}%
\end{equation}
So, while there remains a one-parameter freedom for the PQ charges, corresponding to choices for $\beta$, no such freedom exists for $\mathcal{B}-\mathcal{L}$ charges. 

Second, the ambiguity in the PQ charge of the fermions cannot have any dynamical consequence~\cite{Quevillon:2019zrd, Quevillon:2020hmx}. The simplest way to see this is to adopt a linear representation for the Higgs multiplets. Their couplings with fermions are then uniquely defined, and so are those of the axion.

Third, it is still possible to attain $SU(5)$-invariant PQ charges with the value of $\beta$ quoted in Eq.~(\ref{UnifPQ}). We now understand this value as that for which $\zeta_{Y}^{PQ}=0$, that is, the value which removes $Y$ from the PQ charge in Eq.~(\ref{PQU1U2Y}). Of course, this is compulsory for them to be $SU(5)$ invariant. Yet, it is crucial for the following to realize that at the level of the minimal model, this is nothing more than a choice. In the presence of explicit $\mathcal{B}$ and/or $\mathcal{L}$ violating couplings, other values of $\beta$ may be compulsory.

Finally, the fact that $3\alpha+\beta$ is fixed, see Eq.~(\ref{NSU5fixed}), is
particularly interesting. Looking back at Eq.~(\ref{U1U2charges}), the two
global $U(1)$s have the anomalies%
\begin{equation}
\left(
\begin{array}
[c]{c}%
\partial_{\mu}J_{1}^{\mu}\\
\partial_{\mu}J_{2}^{\mu}%
\end{array}
\right)  =-\frac{N_{f}g_{5}^{2}}{16\pi^{2}}\left(
\begin{array}
[c]{c}%
-1/2\\
+1/2
\end{array}
\right)  A_{\mu\nu}^{A}\tilde{A}^{A,\mu\nu}\ .
\end{equation}
At the low-energy scale, once $SU(5)$ is broken down to the THDM gauge group,
these anomalies can only be matched with that of the PQ current since both
$U(1)_{\mathcal{B}-\mathcal{L}}$ and $U(1)_{Y}$ remain anomaly-free.
Specifically, the anomaly of the PQ current calculated at the $SU(5)$ level,
that is from Eq.~(\ref{PQU1U2Y}) with $\zeta_{1,2}^{PQ}$ in
Eq.~(\ref{PQU1U2Y2}), is
\begin{equation}
\partial_{\mu}J_{PQ}^{\mu}
=-\frac{N_{f}g_{5}^{2}}{16\pi^{2}}\frac{\zeta_{2}^{PQ}-\zeta_{1}^{PQ}}{5} A_{\mu\nu}^{A}\tilde{A}^{A,\mu\nu}
=-\frac{N_{f}g_{5}^{2}}{16\pi^{2}}\mathcal{N}_{SU(5)} A_{\mu\nu}^{A}\tilde{A}^{A,\mu\nu} \ .
\end{equation}
The $\beta$ parameter entering $\zeta_{1,2}^{PQ}$ cancels out, leaving no
parametric freedom. This result matches that computed directly at the level of
the THDM, after the $SU(5)$ breaking. Given the charges in
Eq.~(\ref{PQFinalCharge}), we find%
\begin{equation}
\partial_{\mu}J_{PQ}^{\mu}=\frac{N_{f}}{16\pi^{2}}\left\{  \mathcal{N}%
_{C}g_{s}^{2}G_{\mu\nu}^{a}\tilde{G}^{a,\mu\nu}+\mathcal{N}_{L}g^{2}W_{\mu\nu
}^{i}\tilde{W}^{i,\mu\nu}+\mathcal{N}_{Y}g^{\prime2}B_{\mu\nu}\tilde{B}%
^{\mu\nu}\right\}  \;,
\label{unifano1}
\end{equation}
with, from Eq.~(\ref{JPQTHDM}),
\begin{equation}
\mathcal{N}_{C}=\mathcal{N}_{L}=\frac{3}{5}\mathcal{N}_{Y}=-\mathcal{N}%
_{SU(5)}\ .
\label{unifano2}
\end{equation}
Magically, all the coefficients match, independently of the free parameter $\beta$ in Eq.~(\ref{PQFinalCharge}). The ratio $3/5$ is the usual rescaling factor for the EW-scale hypercharge in terms of the diagonal hypercharge generator of $SU(5)$. Thus, even if there remain some freedom in the PQ
charges, there is no ambiguity in the anomalous coefficients because they have to match that of the $SU(5)$ global anomalies of the $U(1)_{1}\otimes U(1)_{2}$ symmetry (see Ref.~\cite{Ernst:2018bib} for a similar observation in the $SO(10)$ context). Inverting the argument, requiring the matching of the anomalies at the various levels of the symmetry breaking chain provides tight constraints on the fermion PQ charges, though not enough to fix them all.

\subsubsection{Lepton-number violation}

To account for the very light neutrino masses, the standard approach is to implement a seesaw mechanism. In $SU(5)$, as in the SM, this starts by adding a flavor triplet of right-handed neutrinos neutral under the gauge group, $\psi_{\mathbf{1}}=(\nu_{R})^{c}$. Two new couplings are then allowed in the Lagrangian:%
\begin{equation}
\mathcal{L}_{\nu}=-\frac{1}{2}\bar{\psi}_{\mathbf{1}}^{c}%
\mathbf{M}_{R}\psi_{\mathbf{1}}+(\bar{\psi}_{\mathbf{\bar{5}}}^{c%
})_{A}\mathbf{Y}_{1}^{T}\psi_{\mathbf{1}}(h_{i,\mathbf{5}})^{A}%
+h.c.\;,\label{Seesaw}%
\end{equation}
with either $i=1$ or $2$. The important point is that by being neutral, a Majorana mass term is permitted at the GUT scale.

This is where the careful analysis of the $U(1)$ symmetries performed previously pays off. Indeed, if one naively assigns PQ charges to whole $SU(5)$ multiplets, as in Eqs.~(\ref{PQSU5naif1}) and~(\ref{PQSU5naif2}), then no matter the choice of Higgs fiveplet for the neutrino Yukawa coupling, the singlet state cannot be neutral,%
\begin{align}
\label{Majorana1}
\bar{\psi}_{\mathbf{\bar{5}}}^{c}\mathbf{Y}_{1}^{T}\psi_{\mathbf{1}%
}h_{1,\mathbf{5}} &  :PQ(\psi_{\mathbf{1}})=\frac{1}{x}-\frac{3x}{2}\ ,\\
\label{Majorana2} \bar{\psi}_{\mathbf{\bar{5}}}^{c}\mathbf{Y}_{1}^{T}\psi_{\mathbf{1}%
}h_{2,\mathbf{5}} &  :PQ(\psi_{\mathbf{1}})=\frac{2}{x}-\frac{x}{2}\ .
\end{align}
In turn, this would mean that the Majorana mass term breaks the PQ symmetry. So, it may appear that either $\mathbf{M}_{R}$ is forbidden, the axion remains a massless Goldstone boson but there is no seesaw mechanism, or $\mathbf{M}_{R}$ is an explicit PQ-symmetry breaking term but the axion cannot remain massless. It that latter case, it would presumably no longer be able to solve the strong CP puzzle since neutrino masses cannot all be negligible compared to the instanton-induced QCD mass term.

Actually, the axion remains as a Goldstone boson even in the presence of a Majorana mass term because one is not forced to assign $SU(5)$ invariant PQ charges. As detailed in the previous section, in the absence of $\psi
_{\mathbf{1}}$, there was some freedom in how to assign PQ charges. Turning $\mathbf{M}_{R}$ on may use up some of that freedom, but it needs not forbid the PQ symmetry. Let us see how this proceeds in details.

Looking back at the charge assignments in Eq.~(\ref{U1U2charges}), neither $U(1)_{1}$ nor $U(1)_{2}$ survives in the presence of $\mathcal{L}_{\nu_{R}}$. Yet, there is still a global $U(1)$ symmetry active in the Lagrangian, provided the two fiveplets have related charges. Solving for the two possible Yukawa couplings involving the singlet, we find :%
\begin{equation}%
\begin{tabular}
[c]{cccccc}\hline
$U(1)_{W}$ & $h_{1,\mathbf{5}}$ & $h_{2,\mathbf{5}}$ & $\chi_{\mathbf{10}}$ &
$\psi_{\mathbf{\bar{5}}}$ & $\psi_{\mathbf{1}}$\\\hline
$\ \bar{\psi}_{\mathbf{\bar{5}}}^{c}\mathbf{Y}_{1}^{T}\psi
_{\mathbf{1}}h_{1,\mathbf{5}}\ $ & $1$ & $-3/2$ & $-1/2$ & $-1$ & $0$\\
$\bar{\psi}_{\mathbf{\bar{5}}}^{c}\mathbf{Y}_{1}^{T}\psi_{\mathbf{1}%
}h_{2,\mathbf{5}}$ & $1$ & $-1/4$ & $-1/2$ & $1/4$ & $0$\\\hline
\end{tabular}
\label{U1Wcharges}%
\end{equation}
The $U(1)_{W}$ symmetry is spontaneously broken when the fiveplets acquire their vacuum expectation values. In some sense, it replaces the $U(1)_{X}$ symmetry of the minimal $SU(5)$ model. However, and contrary to $U(1)_{X}$, it does lead to a Goldstone boson because there is no global symmetry remaining after the EW symmetry breaking. The $U(1)_{\mathcal{B}-\mathcal{L}} $ is no longer active at the low energy scale because it is broken in the neutrino sector. Thus, the detailed breaking chain is now%
\begin{align}
G_{SU(5)} &  \sim U(1)_{W}\otimes SU(5)\nonumber\\
&  \rightarrow U(1)_{W}\otimes U(1)_{Y}\otimes SU(2)_{L}\otimes SU(3)_{C}%
\nonumber\\
&  \rightarrow U(1)_{em}\otimes SU(3)_{C} \ . 
\label{SSBW}%
\end{align}
There are again two Goldstone bosons from the spontaneous breaking of
$U(1)_{W}\otimes U(1)_{Y}$: one will be the WBG of the $Z$, and the other will
be the axion. Exactly as in the THDM, the physical axion state is defined only
once the fiveplets acquire their VEVs, from its orthogonality to the WBG of
the $Z^{0}$. This again fixes the PQ charges of the $SU(2)_{L}$ doublets in
$h_{1,\mathbf{5}}$ and $h_{2,\mathbf{5}}$ to $x$ and $-1/x$, respectively.

In a way similar to Eq.~(\ref{PQU1U2Y}), the PQ charges can be expressed as
linear combinations of the $U(1)_{Y}$ and $U(1)_{W}$ charges,%
\begin{equation}
PQ=\frac{2}{5}(\xi_{Y}Y+\xi_{W}W)\ .\label{YWcharges}%
\end{equation}
However, contrary to Eq.~(\ref{PQU1U2Y}), it is not possible to have $\xi_{Y}=0$ here since $U(1)_{W}$ is not aligned with $U(1)_{PQ}$. Thus, the PQ charge cannot be $SU(5)$ symmetric.\ The linear combination depends on which fiveplet couples to right-handed neutrinos, and both $\xi_{Y}$ and $\xi_{W}$ depends on the ratio of the vacuum expectation values of the fiveplets since this equation has to reproduce the PQ charges of the doublets $\Phi_{1}\subset h_{1,\mathbf{5}}$ and $\Phi_{2}\subset h_{2,\mathbf{5}}$ given the $W$ charges in Eq.~(\ref{U1Wcharges}). Specifically, by asking in addition that $PQ(\nu_{R})=0$, we find%
\begin{align}
\bar{\psi}_{\mathbf{\bar{5}}}^{c}\mathbf{Y}_{1}^{T}\psi_{\mathbf{1}%
}h_{1,\mathbf{5}} &  :\xi_{Y}=\frac{3}{2}x-\frac{1}{x}\ ,\ \xi_{W}=x+\frac
{1}{x}\ ,\\
\bar{\psi}_{\mathbf{\bar{5}}}^{c}\mathbf{Y}_{1}^{T}\psi_{\mathbf{1}%
}h_{2,\mathbf{5}} &  :\xi_{Y}=\frac{1}{2}x-\frac{2}{x}\ ,\ \xi_{W}=2 \left(
x+\frac{1}{x}\right)  \ .
\end{align}
The PQ charge of the fermions can then be calculated. Without surprise, they correspond to a specific choice for $\beta$ in Eq.~(\ref{PQFinalCharge}):

\begin{subequations}
\label{PQseesaw}%
\begin{align}
\bar{\psi}_{\mathbf{\bar{5}}}^{c}\mathbf{Y}_{1}^{T}\psi_{\mathbf{1}%
}h_{1,\mathbf{5}} &  :PQ(\nu_{R})=\beta+x=0\nonumber\\
&  \Rightarrow PQ(q_{L},u_{R},d_{R},\ell_{L},e_{R})=\left(  -\frac{1}%
{3x},x-\frac{1}{3x},\frac{2}{3x},-x,\frac{1}{x}-x\right)  \ \ ,\\
\bar{\psi}_{\mathbf{\bar{5}}}^{c}\mathbf{Y}_{1}^{T}\psi_{\mathbf{1}%
}h_{2,\mathbf{5}} &  :PQ(\nu_{R})=\beta-\frac{1}{x}=0\nonumber\\
&  \Rightarrow PQ(q_{L},u_{R},d_{R},\ell_{L},e_{R})=\left(  -\frac{x}{3}%
-\frac{2}{3x},-\frac{2x}{3}-\frac{2}{3x},\frac{1}{3x}-\frac{x}{3},\frac{1}%
{x},-\frac{2}{x}\right)  \ \ .
\end{align}
\end{subequations}
This also means that the THDM free parameters $\alpha$ and $\beta$ still satisfy $3\alpha+\beta=2\mathcal{N}_{SU(5)}$, as they should since introducing a gauge singlet does not change the anomaly coefficients. Thus, here also, the anomalies are constant through the SSB chain, and this is the only
manifestation of the underlying $SU(5)$ dynamics on the fermion PQ charges.

\begin{figure}[t]
\centering\includegraphics[width=0.35\textwidth]{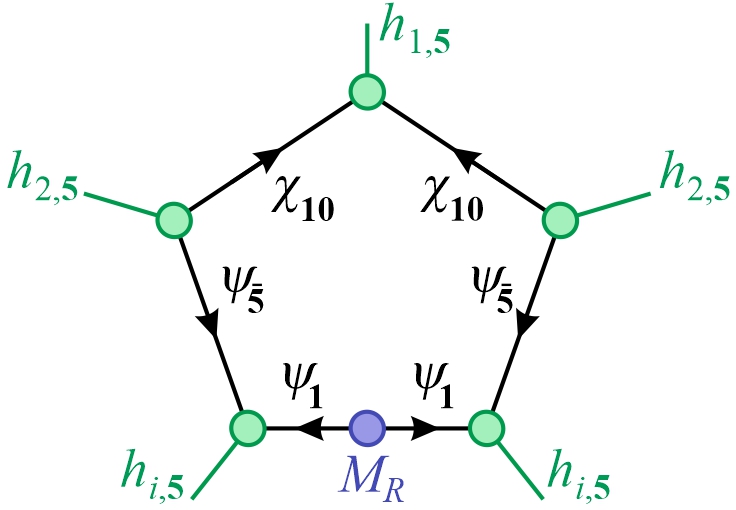}
\caption{Skeleton of the $U(1)_{1}\otimes U(1)_{2}$ symmetry-breaking fermion loop contributing to the effective scalar potential, and corresponding to the effective interactions in Eq.~(\ref{EffScal5h}). Additional derivatives and/or gauge interactions are understood to break the symmetry of the diagram.}%
\label{Fig1}%
\end{figure}

There is still one remaining question to be addressed. The $U(1)_{1}\otimes U(1)_{2}$ symmetry is broken in the fermion sector, leaving only $U(1)_{X}$, but the scalar potential is still invariant under the full $U(1)_{1}\otimes U(1)_{2}$. Only that sector defines the Goldstone boson content of the theory, so one may wonder why the breaking pattern in Eq.~(\ref{SSBW}) is selected and not that in Eq.~(\ref{SSB5}). To understand this, it is necessary to go at the effective potential level. Radiative corrections must break $U(1)_{1}\otimes U(1)_{2}$, leaving only $U(1)_{W}$ as an active symmetry. Further, these radiative corrections must encode the whole fermion sector of the model, since it is the combined presence of the three Yukawa couplings and the Majorana mass term that prevents a $U(1)_{1}\otimes U(1)_{2}$ charge assignment. With this in mind, one can identify the simplest symmetry breaking term as arising from a fermion loop with the generic structure shown in Fig.~\ref{Fig1}, corresponding to
\begin{subequations}
\begin{align}
\bar{\psi}_{\mathbf{\bar{5}}}^{c}\mathbf{Y}_{1}^{T}\psi_{\mathbf{1}%
}h_{1,\mathbf{5}} &  :\mathcal{L}_{scalar}^{eff}=\langle\mathbf{Y}%
_{10}\mathbf{Y}_{5}^{\dagger}\mathbf{Y}_{1}\mathbf{M}_{R}^{-1}\mathbf{Y}%
_{1}\mathbf{Y}_{5}^{\dagger}\rangle\varepsilon_{ABCDE}h_{1,\mathbf{5}}%
^{A}h_{1,\mathbf{5}}^{B}h_{1,\mathbf{5}}^{C}h_{2,\mathbf{5}}^{D}%
h_{2,\mathbf{5}}^{E}+h.c.\ ,\\
\bar{\psi}_{\mathbf{\bar{5}}}^{c}\mathbf{Y}_{1}^{T}\psi_{\mathbf{1}%
}h_{2,\mathbf{5}} &  :\mathcal{L}_{scalar}^{eff}=\langle\mathbf{Y}%
_{10}\mathbf{Y}_{5}^{\dagger}\mathbf{Y}_{1}\mathbf{M}_{R}^{-1}\mathbf{Y}%
_{1}\mathbf{Y}_{5}^{\dagger}\rangle\varepsilon_{ABCDE}h_{1,\mathbf{5}}%
^{A}h_{2,\mathbf{5}}^{B}h_{2,\mathbf{5}}^{C}h_{2,\mathbf{5}}^{D}%
h_{2,\mathbf{5}}^{E}+h.c.\ .
\end{align}
\label{EffScal5h}
\end{subequations}
As such, these couplings vanish due to the antisymmetric contraction, so additional derivatives are understood. Yet, they do break $U(1)_{1}\otimes U(1)_{2}$ but preserve their respective $U(1)_{W}$ symmetry. Also, clearly,
none of them is able to generate a mass term for the pseudoscalar states embedded in the fiveplets, so the WBG of the $Z$ boson and the axion do emerge exactly as in the THDM. One way to see that these couplings can never
contribute to a pseudoscalar mass term is to note that if they could, there would be a way to generate a Majorana mass term for the neutrinos out of the Higgs fields via the diagram obtained by cutting out the Majorana mass term in the fermion loop, see Fig.~\ref{Fig1}. Clearly, this is not possible. Rather, since $\mathcal{B}-\mathcal{L}$ is active in the rest of the $SU(5)$ model, the $\Delta\mathcal{L}=2$ Majorana mass is compensated by a $\Delta\mathcal{B}=2$ combination of colored Higgs states, for example $\varepsilon^{ijk}h_{1,\mathbf{5}}%
^{i}h_{1,\mathbf{5}}^{j}h_{1,\mathbf{5}}^{k}$, and such a combination can never acquire a VEV otherwise QCD would be broken.

\subsubsection{Baryon-number violation}

As we have seen, the ambiguity in the fermion PQ charges can be fixed by adopting a specific neutrino mass model. Once this is done, the PQ symmetry has no more room to accommodate further $\mathcal{B}$ and/or $\mathcal{L}$ violating effects. So, let us briefly investigate what type of violation is allowed, and what happens if we go beyond that.

First, as in the usual $SU(5)$ model, gauge and Higgs states do carry non-trivial $\mathcal{B}-\mathcal{L}$ charges, which are conserved quantum numbers, and allows for $\mathcal{B}+\mathcal{L}$ violating effects. What we want to show now is that the presence of the $U(1)_{1}\otimes U(1)_{2}$ symmetry imposes some restrictions on those effects. To see this, let us start with the dimension-six four-fermion operators
\begin{equation}
\mathcal{H}_{eff}^{\dim6}=\frac{1}{\Lambda^{2}}(c_{1}\ell_{L}q_{L}^{3}%
+c_{2}e_{R}u_{R}^{2}d_{R}+c_{3}e_{R}u_{R}q_{L}^{2}+c_{4}\ell_{L}q_{L}%
d_{R}u_{R})+h.c.\ ,
\end{equation}
where $\Lambda \sim v_{24}$. With $\alpha$ and $\beta$ fixed as in Eq.~(\ref{NSU5fixed}), only the last two are allowed by the PQ symmetry%
\begin{equation}
PQ(e_{R}u_{R}q_{L}^{2}) = PQ(\ell_{L}q_{L}d_{R}u_{R}) = 3\alpha+\beta+\left(  x+\frac{1}{x}\right)  =0 \ .
\label{AllowedOps}
\end{equation}
This is understandable since those two operators can arise from leptoquark gauge interactions, and the PQ symmetry has to be compatible with the gauge structure of the model. Said differently, it is precisely because these operators have to be allowed that $3\alpha+\beta = 2\mathcal{N}_{SU(5)}$. 

On the other hand, the first two operators are forbidden~\cite{Hisano}:
\begin{align}
PQ(\ell_{L}q_{L}^{3}) & = 3\alpha + \beta = -\left(  x+\frac{1}{x}\right)\neq 0  \ ,\\
PQ(e_{R}^{\dagger}u_{R}^{\dagger2}d_{R}^{\dagger}) & = -3\alpha-\beta-2\left(
x+\frac{1}{x}\right)  = -\left(  x+\frac{1}{x}\right) \neq 0 \  .
\end{align}
Basically, the reason for this is the presence of two fiveplets with the Yukawa couplings of Eq.~(\ref{SU5THDM}), which prevents the operator $\bar{\chi}_{\mathbf{10}}^{c}\chi_{\mathbf{10}} \bar{\chi}_{\mathbf{10}}^{c}\psi_{\mathbf{\bar{5}}}$ at tree-level. Instead, operators with that fermionic field content first arise at the dimension-eight level, at which point one can use a quartic Higgs coupling to connect $\bar{\chi}_{\mathbf{10}}^{c}\chi_{\mathbf{10}}$ to $\bar{\chi}_{\mathbf{10}}^{c}\psi_{\mathbf{\bar{5}}}$, so that
\begin{equation}
\mathcal{H}_{eff}^{\dim8}=\frac{1}{\Lambda^{4}}(\bar{\chi}_{\mathbf{10}%
}^{c}\chi_{\mathbf{10}})(\bar{\chi}_{\mathbf{10}}^{c}%
\psi_{\mathbf{\bar{5}}})(h_{1,\mathbf{5}}h_{2,\mathbf{5}}^{\dagger})+...\ .
\end{equation}
The effective operator is then neutral under PQ, as it should. Yet, phenomenologically, these operators end up suppressed by $v_{5}^{2}/\Lambda^{2} \sim v_{5}^{2}/v_{24}^{2}$ compared to the leading dimension-six operators of Eq.~(\ref{AllowedOps}).\footnote{A similar reasoning holds for $\mathcal{B}+\mathcal{L}$ violating operators involving $\psi_{\mathbf{1}}$, which cannot arise from gauge interactions. Depending on which fiveplet enters in the Yukawa coupling $\bar{\psi}_{\mathbf{\bar{5}}}^{c}\mathbf{Y}_{1}^{T}\psi_{\mathbf{1}}$, either $\nu_{R}d_{R}q_{L}^{2}$ or $\nu_{R}u_{R}d_{R}^{2}$ arises at the dimension-six level, the other then being of dimension eight.}

The $SU(5)$ gauge interactions not only break $\mathcal{B}+\mathcal{L}$ via perturbative leptoquark exchanges, but also via non-perturbative instanton interactions. What is peculiar in $SU(5)$ is that these two breaking effects are not aligned: they impose incompatible restrictions on the PQ symmetry. Indeed, while leptoquark interactions ask for $3\alpha+\beta = 2\mathcal{N}_{SU(5)}$, the existence of the instanton effective interactions requires instead $3\alpha+\beta = 0$ since
\begin{equation}
PQ((\chi_{\mathbf{10}}^{3}\psi_{\mathbf{\bar{5}}})^{3}) = 3 (3\alpha + \beta) \ .
\label{SU5inst}
\end{equation}
There is no way to reconcile these constraints since they involve the same combination $3\alpha+\beta$, reflecting the fact that both effects break $\mathcal{B}+\mathcal{L}$ but not $\mathcal{B}-\mathcal{L}$. 

This incompatibility was already indentified in Ref.~\cite{Quevillon:2020hmx}, where it was remarked that the $SU(2)_{L}$ instanton interaction alone, $(\ell_{L}q_{L}^{3})^{3}$, requires to set $3\alpha+\beta = 0$, while the unification of the anomaly coefficients in Eqs.~(\ref{unifano1}) and~(\ref{unifano2}) rather ask for $3\alpha+\beta = 2\mathcal{N}_{SU(5)}$. In this respect, Eq.~(\ref{SU5inst}) provides an interesting new perspective. Indeed, the $SU(5)$ instanton interaction $(\chi_{\mathbf{10}}^{3}\psi_{\mathbf{\bar{5}}})^{3}$ not only contains the electroweak $(\ell_{L}q_{L}^{3})^{3}$ interactions, but also the QCD one, $q_{L}^{6}d_{R}^{\dagger3}u_{R}^{\dagger3}$. This latter interaction must have a non-zero PQ charge since it is another guise of the QCD axial anomaly, $G_{\mu\nu}\tilde{G}^{\mu\nu}$, ultimately responsible for the $\eta^{\prime}$ mass when run down to the low energy scale. So, the very idea of unification requires electroweak instantons to carry non-trivial PQ charges, and one should not expect $3\alpha+\beta = 0$ to hold. 

\begin{figure}[t]
\centering\includegraphics[width=0.40\textwidth]{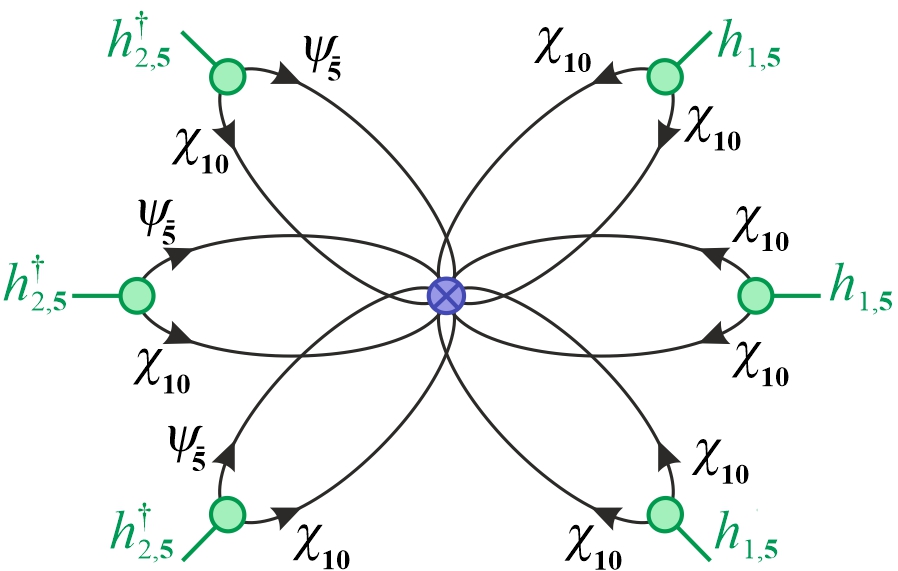}\caption{Effective scalar interaction built on the fermionic instanton interaction of Eq.~(\ref{SU5inst}), and corresponding to the naive estimate of Eq.~(\ref{InstScalar}).}%
\label{Fig2}%
\end{figure}

Phenomenologically, instanton effects are tiny in the perturbative regime. As presented in Ref.~\cite{Quevillon:2020hmx}, a very rough estimate of their impact on the axion mass can be obtained by looking at the effective potential term induced by fermion loops built on the instanton interaction, see Fig.~\ref{Fig2}. The flower-like diagram becomes an effective coupling among Higgs fiveplets:%
\begin{equation}
\mathcal{L}_{scalar}^{eff}\ni\frac{1}{\Lambda^{2}}c_{inst}(h_{\mathbf{5}%
,1}h_{2,\mathbf{5}}^{\dagger})^{3}\ .
\label{InstScalar}
\end{equation}
With $c_{inst}\sim\exp(-4\pi/g^{2})$, the induced axion mass is certainly much smaller than the QCD contribution. Yet, as explained in Ref.~\cite{Quevillon:2020hmx}, this interaction could have cosmological
implications. At high temperature, electroweak instantons may represent the main force driving the alignment of the axion in a specific direction, and the strong CP puzzle would be resolved only at a later stage once hadronic effects start to kick in. 

\subsection{The DFSZ-SU(5) extensions}

As proposed in the original paper, Ref.~\cite{axionDFS}, the axion of the PQ-$SU(5)$ model can be made invisible by following the same recipe as for the THDM. Adding a complex $SU(5)$ singlet $\phi_{\mathbf{1}}$, and allowing for the couplings%
\begin{equation}
V(h_{\mathbf{1}})=\frac{\lambda}{2}(\phi_{\mathbf{1}}^{\dagger}\phi
_{\mathbf{1}})^{2}+\phi_{\mathbf{1}}^{\dagger}\phi_{\mathbf{1}}\left(  \mu
^{2}+\alpha_{1}\langle\mathbf{H}_{\mathbf{24}}^{2}\rangle+\alpha
_{2}h_{1,\mathbf{5}}^{\dagger}h_{1,\mathbf{5}}+\alpha_{3}h_{2,\mathbf{5}%
}^{\dagger}h_{2,\mathbf{5}}\right)  -\left[ \lambda_{12}\phi_{\mathbf{1}%
}^{2}h_{1,\mathbf{5}}^{\dagger}h_{2,\mathbf{5}}+h.c.\right]
\ ,\label{SU5DFSZ}%
\end{equation}
the $U(1)_{1}\otimes U(1)_{2}$ symmetry is preserved provided $\phi_{\mathbf{1}}$ is also charged:
\begin{equation}%
\begin{tabular}
[c]{ccccccc}\hline
& $\phi_{\mathbf{1}}$ & $\mathbf{H}_{\mathbf{24}}$ & $h_{1,\mathbf{5}}$ & $h_{2,\mathbf{5}}$ &
$\chi_{\mathbf{10}}$ & $\psi_{\mathbf{\bar{5}}}$\\\hline
$\ \ U(1)_{1}\ \ $ & $1/2$ & $0$ & $1$ & $0$ & $-1/2$ & $1/2$\\
$\ \ U(1)_{2}\ \ $ & $-1/2$ & $0$ & $0$ & $1$ & $0$ & $1$\\\hline
\label{DFSZSU5}
\end{tabular}
\end{equation}
Another immediate observation from this potential is that even starting with
$\mu^{2}>0$, the GUT symmetry breaking alone is able to induce that of the
$U(1)_{1}\otimes U(1)_{2}$ symmetry whenever $\alpha_{1}<0$. Setting $\mu$ to
zero, the VEV $v_{s}$ of the singlet is automatically set by the GUT scale,
$2\lambda v_{s}^{2}\approx15\alpha_{1}^{2}v_{24}^{2}$.

To proceed, it suffices to plug the polar representations
\begin{equation}
\phi_{\mathbf{1}}=\frac{1}{\sqrt{2}}\exp(i\eta_{s}/v_{s})(v_{s}+\sigma_{s}) \ , \ \  h_{i,\mathbf{5}}=\frac{1}{\sqrt{2}}\exp
(i\eta_{i}/v_{i}) (h_{i,\mathbf{5},1}^{*},h_{i,\mathbf{5},2}^{*},h_{i,\mathbf{5},3}^{*},h_{i,\mathbf{5}}^{+}, v_i + \operatorname{Re} h_{i,\mathbf{5}}^{0}) \ ,
\end{equation}
in the scalar potential, and set all fields but $\eta_{1,2,s}$ to zero. Only
the $-\lambda_{12}\phi_{\mathbf{1}}^{2}h_{1,\mathbf{5}}^{\dagger
}h_{2,\mathbf{5}}$ coupling contributes and%
\begin{equation}
V_{\text{Scalar}}(\eta_{1,2,s})=-\frac{1}{2}\lambda_{12}v_{1}v_{2}v_{s}%
^{2}\cos\left(  \frac{\eta_{1}}{v_{1}}-\frac{\eta_{2}}{v_{2}}-\frac{2\eta_{s}%
}{v_{s}}\right)  \ .
\label{VscalarDFSZ}
\end{equation}
This is exactly the same potential as in the usual THDM version of the DFSZ
model. The mixing matrix is thus
\begin{equation}
\left(
\begin{array}
[c]{c}%
G^{0}\\
a^{0}\\
\pi^{0}%
\end{array}
\right)  =\left(
\begin{array}
[c]{ccc}%
\sin\beta & \cos\beta & 0\\
v_5\cos\beta\sin2\beta/\omega & -v_5\sin\beta\sin2\beta/\omega & v_{s}/\omega\\
v_{s}\cos\beta/\omega & -v_{s}\sin\beta/\omega & -v_5\sin2\beta/\omega
\end{array}
\right)  \left(
\begin{array}
[c]{c}%
\eta_{1}\\
\eta_{2}\\
\eta_{s}%
\end{array}
\right)  \ ,\label{DFSZmixing}%
\end{equation}
with $\omega^{2}=v_{s}^{2}+v_{5}^{2}\sin^{2}2\beta$. The $G^{0}=\sin\beta
\eta_{1}+\cos\beta\eta_{2}$ is the WBG of the $Z^{0}$ boson, the $\pi^{0}$ is the pseudoscalar state occurring in the argument of the cosine function in Eq.~(\ref{VscalarDFSZ}), once properly normalized, and has the mass $M_{\pi^{0}}^{2}\approx\lambda_{12}v_{s}^{2}/\sin2\beta$, while and the axion is orthogonal to both $G^0$ and $\pi^0$, hence
\begin{equation}
a^{0}=\eta_{s}+\frac{v_5}{v_{s}}(\cos\beta\eta_{1}-\sin\beta\eta_{2})\sin
2\beta+\mathcal{O}(v_5^{2}/v_{s}^{2})\ .\label{DFSZaxion}%
\end{equation}
The PQ charges of the scalars can be read off the second line of the mixing matrix in
Eq.~(\ref{DFSZmixing}), and are the same as in Eq.~(\ref{PQdfsz}), while those of the fermions are again given by Eq.~(\ref{PQFinalCharge}).

\subsubsection{Neutrino masses in DFSZ models}

There are many ways to account for neutrino masses in the singlet DFSZ case, so we consider only the simplest realizations: a separate seesaw of type I, the $\nu$DFSZ model, and a seesaw of type II. Let us
briefly describe these three constructions.

\paragraph{Type I seesaw scenario:}

Adding the seesaw terms of Eq.~(\ref{Seesaw}), i.e.,
\begin{equation}
\mathcal{L}_{\nu}=-\frac{1}{2}\bar{\psi}_{\mathbf{1}}^{c}%
\mathbf{M}_{R}\psi_{\mathbf{1}}+(\bar{\psi}_{\mathbf{\bar{5}}}^{c%
})_{A}\mathbf{Y}_{1}^{T}\psi_{\mathbf{1}}(h_{i,\mathbf{5}})^{A}+h.c.\;,
\end{equation}
the active global symmetry is the $U(1)_{W}$ of Eq.~(\ref{U1Wcharges}), under which $\phi_{\mathbf{1}}$ is now charged because of the $\lambda_{12}$ coupling in Eq.~(\ref{SU5DFSZ}):%

\begin{equation}%
\begin{tabular}
[c]{cccccccc}\hline
$U(1)_{W}$ & $\phi_{\mathbf{1}}$ & $h_{1,\mathbf{5}}$ & $h_{2,\mathbf{5}}$ &
$\chi_{\mathbf{10}}$ & $\psi_{\mathbf{\bar{5}}}$ & $\psi_{\mathbf{1}}$ &
$\beta$\\\hline
$\ \bar{\psi}_{\mathbf{\bar{5}}}^{c}\mathbf{Y}_{1}^{T}\psi
_{\mathbf{1}}h_{1,\mathbf{5}}\ $ & $5/4$ & $1$ & $-3/2$ & $-1/2$ & $-1$ & $0$
& $-x$\\
$\bar{\psi}_{\mathbf{\bar{5}}}^{c}\mathbf{Y}_{1}^{T}\psi_{\mathbf{1}%
}h_{2,\mathbf{5}}$ & $5/8$ & $1$ & $-1/4$ & $-1/2$ & $1/4$ & $0$ & $\frac
{1}{x}$\\\hline
\end{tabular}
\label{TypeIU1W}%
\end{equation}
The presence of the singlet does not change the PQ charges of the fermions, still given as in Eq.~(\ref{PQseesaw}). Those charges also correspond to Eq.~(\ref{PQFinalCharge}) with the specific values of $\beta$ quoted in the last column.

\paragraph{$\nu$DFSZ scenario:}

The presence of the singlet opens a new path to generate neutrino masses. The $SU(5)$ version of the $\nu$DFSZ model of Ref.~\cite{Clarke:2015bea}, first discussed in Ref.~\cite{Boucenna:2017fna}, replaces the Majorana mass term in Eq.~(\ref{Seesaw}) by the coupling%
\begin{equation}
\mathcal{L}_{\nu}=-\frac{1}{2}\bar{\psi}_{\mathbf{1}}^{c}%
\mathbf{Y}_{M}\psi_{\mathbf{1}}\phi_{\mathbf{1}}+(\bar{\psi}_{\mathbf{\bar{5}%
}}^{c})_{A}\mathbf{Y}_{1}^{T}\psi_{\mathbf{1}}(h_{i,\mathbf{5}}%
)^{A}+h.c.\;.
\label{LagrnuDFSZ}
\end{equation}
Equivalently, $\phi_{\mathbf{1}}^{\dagger}$ could occur for the first coupling. Altogether, there are thus four possible realization of the $U(1)_{W}$ symmetry. Normalizing $W(h_{1,\mathbf{5}})\equiv1$, the other
fields have the charges%
\begin{equation}%
\begin{tabular}
[c]{ccccccccc}\hline
\multicolumn{2}{c}{$U(1)_{W}$} & $\phi_{\mathbf{1}}$ & $h_{1,\mathbf{5}}$ &
$h_{2,\mathbf{5}}$ & $\chi_{\mathbf{10}}$ & $\psi_{\mathbf{\bar{5}}}$ &
$\psi_{\mathbf{1}}$ & $\beta$\\\hline
$\ \bar{\psi}_{\mathbf{\bar{5}}}^{c}\mathbf{Y}_{1}^{T}\psi
_{\mathbf{1}}h_{1,\mathbf{5}}\ $ & $\bar{\psi}_{\mathbf{1}}^{c%
}\mathbf{Y}_{M}\psi_{\mathbf{1}}\phi_{\mathbf{1}}$ & $1$ & $1$ & $-1$ & $-1/2$
& $-1/2$ & $-1/2$ & $\frac{1}{4x}-\frac{3x}{4}$\\
$\bar{\psi}_{\mathbf{\bar{5}}}^{c}\mathbf{Y}_{1}^{T}\psi_{\mathbf{1}%
}h_{2,\mathbf{5}}$ & $\bar{\psi}_{\mathbf{1}}^{c}\mathbf{Y}_{M}%
\psi_{\mathbf{1}}\phi_{\mathbf{1}}$ & $5/9$ & $1$ & $-1/9$ & $-1/2$ & $7/18$ &
$-5/18$ & $\frac{5}{4x}+\frac{x}{4}$\\
$\bar{\psi}_{\mathbf{\bar{5}}}^{c}\mathbf{Y}_{1}^{T}\psi_{\mathbf{1}%
}h_{1,\mathbf{5}}$ & $\bar{\psi}_{\mathbf{1}}^{c}\mathbf{Y}_{M}%
\psi_{\mathbf{1}}\phi_{\mathbf{1}}^{\dagger}$ & $5/3$ & $1$ & $-7/3$ & $-1/2$
& $-11/6$ & $5/6$ & $-\frac{1}{4x}-\frac{5x}{4}$\\
$\bar{\psi}_{\mathbf{\bar{5}}}^{c}\mathbf{Y}_{1}^{T}\psi_{\mathbf{1}%
}h_{2,\mathbf{5}}$ & $\bar{\psi}_{\mathbf{1}}^{c}\mathbf{Y}_{M}%
\psi_{\mathbf{1}}\phi_{\mathbf{1}}^{\dagger}$ & $5/7$ & $1$ & $-3/7$ & $-1/2$
& $1/14$ & $5/14$ & $\frac{3}{4x}-\frac{x}{4}$\\\hline
\end{tabular}
\label{TablenuDFSZ}%
\end{equation}
Interestingly, the first realization, which is that of Ref.~\cite{Boucenna:2017fna}, is the only one encountered up to now for which $U(1)_{W}$ could be compatible with an underlying $SO(10)$ structure, in
which all the fermions are embedded into a 16 representation of definite $U(1)_{W}$ charge $-1/2$.

At the electroweak scale, the PQ charges of the scalar doublets in $h_{1,\mathbf{5}}$ and $h_{2,\mathbf{5}}$ are still fixed to $x$ and $-1/x$, respectively. For each realization in Eq.~(\ref{TablenuDFSZ}), the PQ symmetry thus corresponds to a specific linear combination of $W$ and $Y$, see Eq.~(\ref{YWcharges}). Once applied to the SM fermions, the PQ charges are unambiguously fixed and correspond to Eq.~(\ref{PQFinalCharge}) with $\beta$ fixed to the values quoted in Eq.~(\ref{TablenuDFSZ}). Note that these charges
are quite complicated, with for example the first realization predicting%
\begin{equation}
PQ(q_{L},u_{R},d_{R},\ell_{L},e_{R},\nu_{R})=\left(  -\frac{x}{12}-\frac
{5}{12x},\frac{11x}{12}-\frac{5}{12x},\frac{7}{12x}-\frac{x}{12},\frac{1}%
{4x}-\frac{3x}{4},\frac{5}{4x}-\frac{3x}{4},\frac{1}{4x}+\frac{x}{4}\right)
\ .
\end{equation}
The fact that the PQ symmetry necessarily arises in part from $U(1)_{Y}$ has completely obscured the underlying $SU(5)$ or $SO(10)$ pattern.

\paragraph{Type II seesaw scenario:}

As a final alternative, let us discuss the $SU(5)$ version of the type II seesaw mechanism~\cite{Dorsner:2005fq}. Instead of right-handed neutrinos, a scalar $\Delta_{\mathbf{15}}$ in the \textbf{15} representation is added, whose branching rule contains an electroweak triplet, $\mathbf{15}\supset(\mathbf{1}\otimes\mathbf{3})_{2}$. This state couples to fermions via%
\begin{equation}
\mathcal{L}_{\nu}=-(\bar{\psi}_{\mathbf{\bar{5}}}^{c})_{A}%
\mathbf{Y}_{\nu}(\psi_{\mathbf{\bar{5}}})_{B}(\Delta_{\mathbf{15}})^{AB}\ ,
\end{equation}
where $(\Delta_{\mathbf{15}})^{AB}=(\Delta_{\mathbf{15}})^{BA}$. In parallel, to entangle the charges of the scalar states, we add to the scalar potential%
\begin{align}
V(\Delta_{\mathbf{15}})  & =\langle\Delta_{\mathbf{15}}^{\dagger}%
\Delta_{\mathbf{15}}\rangle\left(  \mu_{\Delta}^{2}+\beta_{1}\phi_{\mathbf{1}%
}^{\dagger}\phi_{\mathbf{1}}+\beta_{2}\langle\mathbf{H}_{\mathbf{24}}%
^{2}\rangle+\beta_{3}h_{1,\mathbf{5}}^{\dagger}h_{1,\mathbf{5}}+\beta
_{4}h_{2,\mathbf{5}}^{\dagger}h_{2,\mathbf{5}}\right) \nonumber\\
& +\lambda_{15}\langle\Delta_{\mathbf{15}}^{\dagger}\Delta_{\mathbf{15}%
}\rangle^{2}+\lambda_{15}^{\prime}\langle\Delta_{\mathbf{15}}^{\dagger}%
\Delta_{\mathbf{15}}\Delta_{\mathbf{15}}^{\dagger}\Delta_{\mathbf{15}}%
\rangle\nonumber\\
& -\left[  \lambda_{\nu1}\phi_{\mathbf{1}}^{2}h_{1,\mathbf{5}}^{\dagger
}h_{2,\mathbf{5}}+\lambda_{\nu2}\phi_{\mathbf{1}}h_{1,\mathbf{5}}%
h_{2,\mathbf{5}}\Delta_{\mathbf{15}}^{\dagger}+\mu_{\Delta}\lambda_{\nu
3}h_{1,\mathbf{5}}h_{2,\mathbf{5}}\Delta_{\mathbf{15}}^{\dagger}+h.c.\right]
\ .
\end{align}
Because of the last two couplings in the last line, which becomes $\Delta_{\mathbf{15}}$ tadpoles after the symmetry breaking, that field develops a small VEV even when the terms in the first line sum up to a large and positive mass term. When $\beta_{2}\sim \mathcal{O}(1)$ and $v_{24}>v_{s},\mu_{\Delta}$,
\begin{equation}
v_{15}\sim\frac{v_{1}v_{2}}{\beta_{2}v_{24}^{2}}\left(  v_{s}\lambda_{\nu
2}+\mu_{\Delta}\lambda_{\nu3}\right)  \ .
\end{equation}
As detailed in Ref.~\cite{Quevillon:2020hmx}, if the three $\lambda_{\nu i}$ couplings are present, there is no global symmetry and no axion. If $\lambda_{\nu1}=0$, then there is a global symmetry but $\phi_{\mathbf{1}}$ is neutral and decouples. This means that the axion is embedded into the $h_{1,\mathbf{5}}$, $h_{2,\mathbf{5}}$,
with a small $\Delta_{\mathbf{15}}$ component, and couples too strongly to SM particles. The two viable DFSZ-like scenarios are those with either $\lambda_{\nu2}=0$ or $\lambda_{\nu3}=0$, and this latter scenario further exists under two incarnations depending on whether $\phi_{\mathbf{1}}$ or $\phi_{\mathbf{1}}^{\dagger}$ occurs in the $\lambda_{\nu2}$ coupling. For those three scenarios, there is enough room for a global $U(1)_{W}$ symmetry, with charges%
\begin{equation}%
\begin{tabular}
[c]{ccccccccc}\hline
\multicolumn{2}{c}{$U(1)_{W}$} & $\phi_{\mathbf{1}}$ & $\Delta_{\mathbf{15}}$
& $h_{1,\mathbf{5}}$ & $h_{2,\mathbf{5}}$ & $\chi_{\mathbf{10}}$ &
$\psi_{\mathbf{\bar{5}}}$ & $\beta$\\\hline
$\phi_{\mathbf{1}}^{2}h_{1,\mathbf{5}}^{\dagger}h_{2,\mathbf{5}}$ &
$\phi_{\mathbf{1}}h_{1,\mathbf{5}}h_{2,\mathbf{5}}\Delta_{\mathbf{15}%
}^{\dagger} $ & $1$ & $1$ & $1$ & $-1$ & $-1/2$ & $-1/2$ & $\frac{1}{4x}%
-\frac{3x}{4}$\\
$\phi_{\mathbf{1}}^{2}h_{1,\mathbf{5}}^{\dagger}h_{2,\mathbf{5}}\ $ &
$\phi_{\mathbf{1}}^{\dagger}h_{1,\mathbf{5}}h_{2,\mathbf{5}}\Delta
_{\mathbf{15}}^{\dagger}$ & $5/7$ & $-1/7$ & $1$ & $-3/7$ & $-1/2$ & $1/14$ &
$\frac{3}{4x}-\frac{x}{4}$\\
$\phi_{\mathbf{1}}^{2}h_{1,\mathbf{5}}^{\dagger}h_{2,\mathbf{5}}$ &
$h_{1,\mathbf{5}}h_{2,\mathbf{5}}\Delta_{\mathbf{15}}^{\dagger}$ & $5/6$ &
$1/3 $ & $1$ & $-2/3$ & $-1/2$ & $-1/6$ & $\frac{1}{2x}-\frac{x}{2}$\\\hline
\end{tabular}
\label{TypeIIU1W}%
\end{equation}
From this, the PQ charge are trickier to find because part of $\operatorname{Im}\Delta_{\mathbf{15}}^{55}$ is eaten by the $Z^{0}$ boson, and the axion has to be orthogonal to that state. However, we showed in Ref.~\cite{Quevillon:2020hmx} that to leading order in $v_{15}/v_{5}$, the PQ charge of $h_{1,\mathbf{5}}$
and $h_{2,\mathbf{5}}$ do remain equal to $x$ and $-1/x$. With this information, the pattern of fermion charges is found to match again Eq.~(\ref{PQFinalCharge}), up to negligible $\mathcal{O}(v_{15}/v_{5})$ corrections, with $\beta$ given in the last column of Eq.~(\ref{TypeIIU1W}). Note that the corresponding PQ charges of the weak triplet in $\Delta_{\mathbf{15}}$ for the three scenarios are $3x/2-1/2x$, $x/2-3/2x$, and $x-1/x$, respectively, in agreement with Ref.~\cite{Quevillon:2020hmx}.

\subsubsection{The adjoint DFSZ model}

In the previous section, the DFSZ extension was performed with the help of a complex singlet field, whose presence has no other motivation than moving the axion scale at the GUT scale. A somewhat simpler model, first proposed in Ref.~\cite{Wise:1981ry} is obtained by instead using the scalar fields already present at the GUT scale, that is, the $\mathbf{H}_{\mathbf{24}}$. Indeed, by making this field complex, the $Z_{2}$ symmetry of the minimal model is extended to a continuous global $U(1)_{24}$ symmetry, corresponding to the phase redefinitions $\mathbf{H}_{\mathbf{24}}\rightarrow\exp(i\alpha)\mathbf{H}_{\mathbf{24}}$.

When $\mathbf{H}_{\mathbf{24}}$ acquires its VEV, the $U(1)_{24}$ symmetry is spontaneously broken at the GUT scale, and an additional Goldstone boson arises. For this field to be the axion, we must force the $U(1)_{24}$ symmetry to be anomalous, which requires fermions to transform non-trivially under $U(1)_{\mathbf{24}}$. Since $\mathbf{H}_{\mathbf{24}}$ does not directly couple to fermions, this has to proceed through charging the Higgs fiveplets under $U(1)_{24}$ first. It is at this stage that two fiveplets are required. Indeed, putting back the $SU(5)$ indices, the $\mathbf{H}_{\mathbf{24}}$ can only communicate its charge via $(\mathbf{H}_{\mathbf{24}})_{B}^{A}$ or $(\mathbf{H}_{\mathbf{24}})_{C}^{A}(\mathbf{H}_{\mathbf{24}})_{B}^{C}$. With only one fiveplet, contracting the indices necessarily requires an equal number of $(h_{\mathbf{5}})^{B}$ and $(h_{\mathbf{5}}^{\dagger})_{A}$, and these couplings end up forbidden by the $U(1)_{24}$ symmetry. With
two fiveplets, on the other hand, the contractions%
\begin{equation}
V_{scalar}(h_{1,\mathbf{5}},h_{2,\mathbf{5}},\mathbf{H}_{\mathbf{24}}%
)\supset\gamma_{1}(h_{1,\mathbf{5}}^{\dagger}h_{2,\mathbf{5}})\langle
\mathbf{H}_{\mathbf{24}}\mathbf{H}_{\mathbf{24}}\rangle+\gamma_{2}%
(h_{1,\mathbf{5}}^{\dagger}\mathbf{H}_{\mathbf{24}}\mathbf{H}_{\mathbf{24}%
}h_{2,\mathbf{5}})+h.c.\ ,\label{SU5coupl}%
\end{equation}
are allowed provided $h_{1,\mathbf{5}}$ and $h_{2,\mathbf{5}}$ have different
$U(1)_{24}$ charges. So, the symmetry of this model is again $U(1)_{1}\otimes
U(1)_{2}\sim U(1)_{X}\otimes U(1)_{24}$, with $\mathbf{H}_{\mathbf{24}}$
charged under both $U(1)_{1}$ and $U(1)_{2}$:
\begin{equation}%
\begin{tabular}
[c]{cccccc}\hline
& $\mathbf{H}_{\mathbf{24}}$ & $h_{1,\mathbf{5}}$ & $h_{2,\mathbf{5}}$ &
$\chi_{\mathbf{10}}$ & $\psi_{\mathbf{\bar{5}}}$\\\hline
$\ \ U(1)_{1}\ \ $ & $1/2$ & $1$ & $0$ & $-1/2$ & $1/2$\\
$\ \ U(1)_{2}\ \ $ & $-1/2$ & $0$ & $1$ & $0$ & $1$\\\hline
\end{tabular}
\label{H24sym}%
\end{equation}
At leading order in $v_{5}/v_{24}$, the axion is entirely embedded in $\mathbf{H}_{\mathbf{24}}$ and does not couple to fermions. Indeed, at the EW level, there is no remaining symmetry beyond the local SM gauge symmetries, hence no new Goldstone boson (besides that eaten by the $Z^{0}$). This is evident since $(h_{1,\mathbf{5}}^{\dagger})_{A}(\mathbf{v}_{\mathbf{24}})_{C}^{A}(\mathbf{v}_{\mathbf{24}})_{B}^{C}(h_{2,\mathbf{5}})^{B}$ breaks the $U(1)_{1}\otimes U(1)_{2}$ symmetry explicitly. But, as in the DFSZ model, this picture gets modified at $\mathcal{O}(v_{5}/v_{24})$, once both breaking stages are combined, since the coupling Eq.~(\ref{SU5coupl}) is also an explicit $U(1)_{24}$ breaking term when $h_{i,\mathbf{5}}$ acquire their vacuum expectation values. Thus, the true Goldstone boson arising at the low scale is a combination of the pseudoscalar $\operatorname{Im}\mathbf{H}_{\mathbf{24}}^{0}$ and $\operatorname{Im}h_{i,\mathbf{5}}^{0}$, with the latter $\operatorname{Im}h_{i,\mathbf{5}}$ components suppressed by $v_{i}/v_{24}$.

The most general scalar potential involves many couplings because $\mathbf{H}_{\mathbf{24}}^{\dagger}$ and $\mathbf{H}_{\mathbf{24}}$ transform in the same way, and will not be written down explicitly. Let us assume it supports the breaking chain of Eq.~(\ref{SSB5}). To identify the pseudoscalar states, the simplest is to use the polar representations%
\begin{equation}
\mathbf{H}_{\mathbf{24}}=\frac{1}{\sqrt{2}}\exp(i\eta_{24}/v_{24}%
)\mathbf{v}_{\mathbf{24}}+...\ ,\ h_{i,\mathbf{5}}=\frac{1}{\sqrt{2}}%
\exp(i\eta_{i}/v_{i})\mathbf{v}_{i,\mathbf{5}}+...\ ,
\end{equation}
with $\mathbf{v}_{\mathbf{24}}=v_{24}\operatorname{diag}(1,1,1,-(3+\varepsilon)/2,-(3-\varepsilon)/2)$ and $\mathbf{v}_{i,\mathbf{5}}=(0,0,0,0,v_{i})^{T}$, and we have not written the other states explicitly. Plugging this in the potential, and restricted to the $\eta_{i}$ fields, only the couplings in Eq.~(\ref{SU5coupl}) survive because all the rest is invariant under $U(1)_{1}\otimes U(1)_{2}\otimes U(1)_{24}$, and the $\eta_{i}$ fields cancel out. Thus,%
\begin{equation}
V_{scalar}(\eta_{1},\eta_{2},\eta_{24})=\frac{1}{8}v_{1}v_{2}v_{24}^{2}\left(
2\gamma_{1}(15+\varepsilon^{2})+\gamma_{2}(3-\varepsilon)^{2}\right)
\cos\left(  \frac{\eta_{1}}{v_{1}}-\frac{\eta_{2}}{v_{2}}-\frac{2\eta_{24}%
}{v_{24}}\right)  \ .
\end{equation}
Apart from the prefactor, the cosine dependence on the pseudoscalars is exactly the same as in the DFSZ model. Thus, the mixing matrix is that in Eq.~(\ref{DFSZmixing}), with $v_{s}\rightarrow v_{24}$, and the pattern of PQ charge is not modified. For instance, scalars have the charges%
\begin{equation}
PQ(\Phi_{1}\subset h_{1,\mathbf{5}},\Phi_{2}\subset h_{2,\mathbf{5}}%
,\phi\subset \mathbf{H}_{\mathbf{24}})=\left(  x\ \ ,\ -\frac{1}{x}\ ,\ \frac{1}{2}\left(
x+\frac{1}{x}\right)  \right)  \ \ ,
\label{PQaDFSZ}
\end{equation}
where $\phi$ is the extra electroweak singlet arising when $\mathbf{H}_{\mathbf{24}}$ is made complex. With this, the SM fermions retain their PQ charges as given in Eq.~(\ref{PQFinalCharge}), including the one-parameter ambiguity. Note that if we assign to the singlet in $\mathbf{H}_{\mathbf{24}}$ twice the above PQ charge by replacing the $\gamma_{1,2}$ couplings of Eq.~(\ref{SU5coupl}) by $\gamma(h_{1,\mathbf{5}%
}^{\dagger}\mathbf{H}_{\mathbf{24}}h_{2,\mathbf{5}})$, the PQ charges of the doublets and SM fermions stay the same.

Concerning neutrinos, the seesaw type I (with right-handed neutrinos $\psi_{1}$) and type II (with scalars $\Delta_{15}$) can again be constructed, with the $U(1)_{W}$ symmetry identified as in Eq.~(\ref{TypeIU1W}) and~(\ref{TypeIIU1W}), with $\phi_{\mathbf{1}}\rightarrow\mathbf{H}_{\mathbf{24}}$. The final PQ charges obviously stay the same\footnote{In this respect, it should be mentioned that the adjoint DFSZ model, with and without the type I seesaw, has been presented in Ref.~\cite{FileviezPerez:2019ssf}, but the precise identification of the symmetry breaking chain, including the entanglement with $U(1)_{\mathcal{B}-\mathcal{L}}$, and the actual PQ charges were not discussed there.}. The only difference is that no equivalent of the $\nu$DFSZ model exists with an adjoint field, since there is no way to couple $\mathbf{H}_{\mathbf{24}}$ directly to a pair of right-handed neutrinos.

\section{Flipped SU(5) axion models\label{sec4}}

In the minimal $SU(5)$ model, $\mathcal{B}-\mathcal{L}$ emerges from the global $U(1)_{X}$ symmetry. Because a global $U(1)$ is active throughout the SSB chain, we ended up with a one-parameter freedom in the definition of the PQ charges of the fermions. This freedom was then used to make the PQ solution compatible with various seesaw mechanisms, which explicitly break the global symmetry.

Since the $U(1)_{X}$ symmetry is not anomalous, nothing prevents it from becoming local. This opens an alternative realization of $SU(5)$ because the SM hypercharge $U(1)_{Y}$ need not be entirely generated by the generator $T_{24}$ of $SU(5)$. Instead, it could emerge as an unbroken linear combination of $U(1)_{X}$ and $U(1)_{Y^{\prime}}\subset SU(5)$. This is called the flipped $SU(5)$ model~\cite{Barr:1981qv,Derendinger:1983aj,Antoniadis:1987dx, Ellis:1988tx}. Because the $X$ symmetry, and therefore also $\mathcal{B}-\mathcal{L}$, is realized differently, our goal is to analyze the consequences on the PQ or DFSZ mechanism, once embedded in this framework.

\subsection{Brief overview of the flipped SU(5) model}

Before entering into the discussion of axionic aspects, let us briefly summarize the construction of the flipped model. Starting with $Y=\alpha Y^{\prime}+\beta X$, and requiring that the fermion charges sum up to the appropriate $Q=T^{3}+Y/2$, there are only two solutions for $\alpha$ and $\beta$. The minimal model with $\alpha=1$, $\beta=0$, and the flipped one with $\alpha=-1/5$ and $\beta=2/5$. In that latter case, the fermions are embedded into $U(1)_{X}\otimes SU(5)$ multiplets as
\begin{subequations}
\label{FlippedF}%
\begin{align}
(\mathbf{1})_{5} &  =(\mathbf{1}\otimes\mathbf{1})_{2}\Rightarrow
\chi_{\mathbf{1}}=e_{R}^{-c}\ ,\\
(\mathbf{5}^{\ast})_{-3} &  =(\mathbf{\bar{3}}\otimes\mathbf{1})_{-4/3}%
\oplus(\mathbf{1}\otimes\mathbf{2})_{-1}\Rightarrow\psi_{\mathbf{\bar{5}}%
}=u_{R}^{c}\oplus\ell_{L}\ ,\\
(\mathbf{10})_{1} &  =(\mathbf{\bar{3}}\otimes\mathbf{1})_{2/3}\oplus
(\mathbf{3}\otimes\mathbf{2})_{1/3}\oplus(\mathbf{1}\otimes\mathbf{1}%
)_{0}\Rightarrow\chi_{\mathbf{10}}=d_{R}^{c}\oplus q_{L}\oplus\nu
_{R}^{c}\ .
\end{align}
Altogether, the fermion representations are thus obtained from those of the minimal model by flipping the right-handed components, $e_{R}^{-c}\leftrightarrow\nu_{R}^{c}$ and $u_{R}^{c}\leftrightarrow d_{R}^{c}$, and leaving the left-handed components in place.

The $X$ charges are fixed up to an overall normalization. To parametrize this freedom, we write the $SU(5)\otimes U(1)_{X}$ derivative as
\end{subequations}
\begin{equation}
D^{\mu}=\partial^{\mu}+i\sqrt{\frac{3}{5}}g_{5}\frac{Y^{\prime}}{2}%
B^{\prime\mu}+ig_{X}\frac{X}{2}X^{\mu}+... \ , 
\label{CovFlip}%
\end{equation}
where the $SU(5)$ gauge boson associated with $U(1)_{Y^{\prime}}$ is denoted $B_{\mu}^{\prime}$, while that of $U(1)_{X}$ by $X_{\mu}$. In the minimal case, $\sqrt{5/3}g^{\prime}=g_{5}=g$, from which $\sin^{2}\theta_{W}=3/8$. In the flipped case, this relation is altered because $U(1)_{Y}$ has two uncorrelated origins and the normalization of the $g_{X}$ relative to $g_{5}$ is free. Fixing the $X$ charges as in Eq.~(\ref{FlippedF}), the normalization freedom resurfaces in the coupling constants, and we write $g_{X}=Kg_{5}$ for some constant $K$. The SM field $B^{\mu}$ is a linear combination of $X^{\mu}$
and $B^{\prime\mu}$, say $X_{\mu}=\sin\theta_{X}B_{\mu}^{X}+\cos\theta_{X}B_{\mu}$ and $B^{\prime\mu}=\cos\theta_{X}B_{\mu}^{X}-\sin\theta_{X}B_{\mu}$. Plugging this in the covariant derivative Eq.~(\ref{CovFlip}), and requiring that it collapses to $D^{\mu}=\partial^{\mu}+ig^{\prime}(Y/2)B^{\mu
}+...$ with $Y=\alpha Y^{\prime}+\beta X$ gives
\begin{equation}
\tan\theta_{X}=\sqrt{5/12}K\ \ ,\ \ \ 15g_{5}^{2}=\left(  1+\frac{12}{5K^{2}%
}\right)  g^{\prime2}\ ,\label{FlipGaugeC2}%
\end{equation}
where we have set $\alpha=-1/5$ and $\beta=2/5$. With this, and since $g_{5}=g$ at the GUT scale as $SU(2)_{L}\subset SU(5)$, the prediction for $\sin^{2}\theta_{W}$ is modified to%
\begin{equation}
\sin^{2}\theta_{W}\equiv\frac{3}{8}\frac{50K^{2}}{20K^{2}+3}\ .
\end{equation}

Originally, the idea of $SU(5)\otimes U(1)_{X}$ was that it could emerge from $SO(10)$. In that case, $g_{X}=g_{5}=g_{10}$ at the unification scale, but the $X$ charges are not normalized to $X_{\mathbf{10}}=1$, $X_{\mathbf{\bar{5}}}=-3$, and $X_{\mathbf{1}}=5$. To find the correct relative normalization
under the assumption that $SO(10)\rightarrow SU(5)\otimes U(1)_{X}$, one can use the fact that both $Y^{\prime2}$ and $X^{2}$ must sum up to the same over a complete $SO(10)$ representation. Since $\mathbf{1}\oplus\mathbf{5}^{\ast}\oplus\mathbf{10}=\mathbf{16}$, and with the $X$ charge of
Eq.~(\ref{FlippedF}), we find $K^{2}=1/10$ and $\sin^{2}\theta_{W}=3/8$, precisely as in the minimal model.

To achieve $SU(5)\otimes U(1)_{X}\rightarrow SU(3)_{C}\otimes SU(2)_{L}\otimes U(1)_{Y}$, both $SU(5)$ and $U(1)_{X}$ must be broken simultaneously at the GUT scale. This can be achieved only by giving a vacuum expectation value to a state with $Y^{\prime}\neq0$, $X\neq0$ but $Y=0$. The usual $\mathbf{H}_{\mathbf{24}}$ cannot be used since all its states have $Y^{\prime}=0$. Instead, the simplest representations able to induce a consistent symmetry breaking are a $\mathbf{H}_{\mathbf{10}}$ with $X_{\mathbf{10}}=+1$ or a
$\mathbf{H}_{\mathbf{50}}$ with $X_{\mathbf{50}}=-2$, since 
\begin{align}
\mathbf{10}_{1} &  =(\mathbf{\bar{3}}\otimes\mathbf{1})_{2/3}\oplus
(\mathbf{3}\otimes\mathbf{2})_{1/3}\oplus(\mathbf{1}\otimes\mathbf{1}%
)_{0}\;,\\
\mathbf{50}_{-2} &  =(\mathbf{8}\otimes\mathbf{2})_{-1}\oplus(\mathbf{\bar{6}%
}\otimes\mathbf{3})_{-2/3}\oplus(\mathbf{6}\otimes\mathbf{1})_{-4/3}%
\nonumber\\
&  \ \ \ \ \oplus(\mathbf{\bar{3}}\otimes\mathbf{2})_{-1/3}\oplus
(\mathbf{3}\otimes\mathbf{1})_{-2/3}\oplus(\mathbf{1}\otimes\mathbf{1})_{0}\;.
\end{align}
The symmetry breaking pattern is the same in both cases, the only difference being $M_{B_{X}}^{2}/M_{X,Y}^{2}=K^{2}+12/5$ when using the $\mathbf{H}_{\mathbf{10}}$, and twice that using the $\mathbf{H}_{\mathbf{50}}$. 

For the electroweak symmetry breaking, if only one multiplet is introduced, it must transform as a fiveplet with charge $X_{\mathbf{5}}=-2$. Indeed, once $e_{R}$ is a singlet, lepton masses can only come from a term $\bar{\psi}_{\mathbf{\bar{5}}}^{c}\psi_{\mathbf{1}}h$, thus $h$ must transform as a fiveplet.
The states are
\begin{equation}
(\mathbf{5})_{-2}=(\mathbf{3}\otimes\mathbf{1})_{-2/3}\oplus(\mathbf{1}%
\otimes\mathbf{2})_{-1}\Rightarrow h_{\mathbf{5}}=h_{i}^{\ast}\oplus\left(
\begin{array}
[c]{c}%
h^{0\ast}\\
-h^{-}%
\end{array}
\right)  \ .
\end{equation}
Notice that compared to the fermions in $\psi_{\mathbf{\bar{5}}}$, the charged conjugate $SU(2)_{L}$ spinor appears, and the colored $h_{i}^{\ast}$ states end up being the same as in the minimal $SU(5)$ model.

The most general potential involving $h_{\mathbf{5}}$ and $\mathbf{H}_{\mathbf{10}}$ is%
\begin{align}
V(h_{\mathbf{5}},\mathbf{H}_{\mathbf{10}}) &  =-\frac{\mu^{2}}{2}%
\langle\mathbf{H}_{\mathbf{10}}^{\dagger}\mathbf{H}_{\mathbf{10}}\rangle
+\frac{a}{4}\langle\mathbf{H}_{\mathbf{10}}^{\dagger}\mathbf{H}_{\mathbf{10}%
}\rangle^{2}+\frac{b}{2}\langle\mathbf{H}_{\mathbf{10}}^{\dagger}%
\mathbf{H}_{\mathbf{10}}\mathbf{H}_{\mathbf{10}}^{\dagger}\mathbf{H}%
_{\mathbf{10}}\rangle\nonumber\\
&  \ \ \ \ -\frac{\nu^{2}}{2}(h_{\mathbf{5}}^{\dagger}h_{\mathbf{5}}%
)+\frac{\lambda}{4}(h_{\mathbf{5}}^{\dagger}h_{\mathbf{5}})^{2}+\alpha
(h_{\mathbf{5}}^{\dagger}h_{\mathbf{5}})\langle\mathbf{H}_{\mathbf{10}%
}^{\dagger}\mathbf{H}_{\mathbf{10}}\rangle+\beta h_{\mathbf{5}}^{\dagger
}\mathbf{H}_{\mathbf{10}}^{\dagger}\mathbf{H}_{\mathbf{10}}h_{\mathbf{5}%
}\nonumber\\
&  \ \ \ \ +\left(  \gamma\varepsilon_{ABCDE}\mathbf{H}_{\mathbf{10}}%
^{AB}\mathbf{H}_{\mathbf{10}}^{CD}h_{\mathbf{5}}^{E}+h.c.\right)
\;.\label{PotH10H10h5}%
\end{align}
We will not analyze this potential in details, but comment on a few saillant features. First, in a combined treatment of the GUT and electroweak symmetry breaking, there is no modification of $\langle0|\mathbf{H}_{\mathbf{10}}|0\rangle$ since it transforms as an electroweak singlet. Second, the
antisymmetric $\gamma$ coupling does not play any role in the symmetry breaking, and vanishes at the minimum. However, it directly couples the colored components $(\mathbf{3}\otimes\mathbf{1})_{-2/3}\subset h_{\mathbf{5}%
}$ to that of the $\mathbf{H}_{\mathbf{10}}$. With $\gamma$ of $\mathcal{O}(1)$, these fields naturally end up at the GUT scale. This alleviates the doublet-triplet problem of the minimal model, since there is here no need to fine-tune the $\alpha$ and $\beta$ couplings in Eq.~(\ref{PotH10H10h5}) to make the $h_{i}^{\ast}$ heavy enough while maintaining $\langle0|h_{\mathbf{5}}|0\rangle$ at the electroweak scale. The situation with $\mathbf{H}_{\mathbf{50}}$ is similar, though with a quartic term $\varepsilon^{ABGHI}(\mathbf{H}_{\mathbf{50}})_{ABCD}(\mathbf{H}_{\mathbf{50}}^{\dagger})^{CDEF}(\mathbf{H}_{\mathbf{50}})_{EFGH}(h_{\mathbf{5}}^{\dagger})_{I}$ in place of the cubic $\varepsilon_{ABCDE}\mathbf{H}_{\mathbf{10}}^{AB}\mathbf{H}_{\mathbf{10}}^{CD}h_{\mathbf{5}}^{E}$ interaction.

Given the $U(1)_{X}$ charges of the fermions and of the electroweak Higgs boson $h_{\mathbf{5}}$, the possible Yukawa couplings are the same as in the minimal model:
\begin{align}
\mathcal{L}_{\text{Yukawa}}  & =-\frac{1}{4}\varepsilon_{ABCDE}(\bar{\chi
}_{\mathbf{10}}^{c})^{AB}\mathbf{Y}_{10}(\chi_{\mathbf{10}}%
)^{CD}h_{\mathbf{5}}^{E}+\sqrt{2}(\bar{\psi}_{\mathbf{\bar{5}}}^{c%
})_{A}\mathbf{Y}_{5}(\chi_{\mathbf{10}})^{AB}(h_{\mathbf{5}}^{\dagger}%
)_{B}\nonumber\\
& \ \ \ \ +(\bar{\psi}_{\mathbf{\bar{5}}}^{c})_{A}\mathbf{Y}_{1}^{T}%
\psi_{\mathbf{1}}(h_{\mathbf{5}})^{A}+h.c.\;.\label{YukFlip}%
\end{align}
But since fermions are flipped, the mass relation of the minimal model is
replaced by%
\begin{equation}
\mathbf{Y}_{d}=\mathbf{Y}_{10}=\mathbf{Y}_{10}^{T}\;,\;\;\mathbf{Y}%
_{u}=\mathbf{Y}_{\nu}^{T}=\mathbf{Y}_{5}\;,\ \ \mathbf{Y}_{e}=\mathbf{Y}%
_{1}\ .\label{MassFlip}%
\end{equation}
The annoying relationship between $m_{d}$ and $m_{e}$ is relaxed, but at the cost of a totally inconsistent neutrino sector. Contrary to the situation in the minimal model, neutrino masses must be addressed for the flipped construction to make sense.

Actually, all the necessary ingredients are already present to automatically solve the neutrino problem. As in the minimal model, the gauge dynamics breaks $\mathcal{B}+\mathcal{L}$. However, since $U(1)_{X}$ is also spontaneously broken, neither $\mathcal{B}$ nor $\mathcal{L}$ survives below the GUT scale. Looking back at Eq.~(\ref{FlippedF}), it is evident that the state $(\mathbf{1}\otimes\mathbf{1})_{0}\subset\mathbf{H}_{\mathbf{10}}$ developing a vacuum expectation value carries $\mathcal{L}=1$. Similarly, the state $(\mathbf{1}\otimes\mathbf{1})_{0}\subset\mathbf{H}_{\mathbf{50}} $ carries $\mathcal{L}=-2$, as can be guessed from $\mathbf{10}_{-1}^{\ast}\otimes\mathbf{10}_{-1}^{\ast}=\mathbf{50}_{-2}\oplus\mathbf{45}_{-2}%
\oplus\mathbf{5}_{-2}$. Thus, the symmetry breaking at the GUT scale can induce $\Delta\mathcal{L}$ effects, and in particular, it can be used to generate a Majorana mass for the neutrinos.

With the $H_{\mathbf{50}}$, this is trivial to achieve since we should actually include in $\mathcal{L}_{\text{Yukawa}}$ the coupling%
\begin{equation}
\frac{1}{2}(\bar{\chi}_{\mathbf{10}}^{c})^{AB}Y_{\mathbf{50}}%
(\chi_{\mathbf{10}})^{CD}(\mathbf{H}_{\mathbf{50}})_{AB,CD}\rightarrow
-\frac{v_{50}}{\sqrt{2}}\bar{\nu}_{R}^{c}\nu_{R}\ .\label{Majo50}%
\end{equation}
From there, the seesaw mechanism proceeds as usual, and neutrino masses end up being proportional to%
\begin{equation}
m_{\nu}=\frac{v_{5}^{2}}{v_{50}}\mathbf{Y}_{u}^{T}(\mathbf{Y}_{\mathbf{50}%
})^{-1}\mathbf{Y}_{u}\ .
\end{equation}
Note that $v_{5}^{2}/v_{50}\sim10^{-2}~$eV when $v_{50}$ is of $\mathcal{O}(10^{16}$ GeV$)$, thus $\mathbf{Y}_{\mathbf{50}}$ needs to be somewhat suppressed to ensure one neutrino states is heavy enough to account for the atmospheric mass splitting $\Delta m_{atm}^{2}\approx2.5\times10^{-3}\,$eV$^{2}$. 

\begin{figure}[t]
\centering\includegraphics[width=0.32\textwidth]{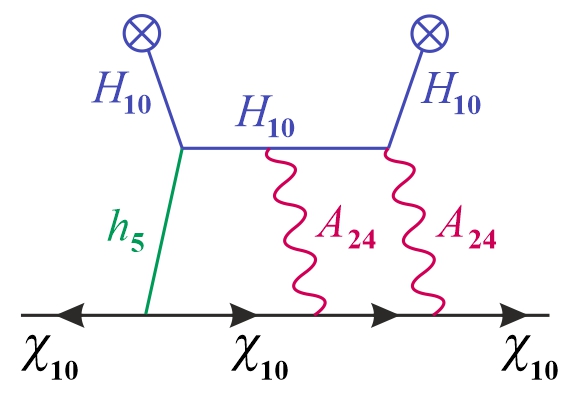}
\caption{The mechanism of Ref.~~\cite{Witten:1979nr} adapted to the flipped $SU(5)$ model~\cite{Rodriguez:2013rma}, leading to a GUT-scale Majorana mass term for $\nu_R$ when $\mathbf{H}_{\mathbf{10}}$ acquires its VEV, see Eqs.~(\ref{MajoH10}) and~(\ref{MajoH10b}). This diagram exists thanks to the cubic interaction $\gamma\mathbf{H}_{\mathbf{10}}\mathbf{H}_{\mathbf{10}}h_{\mathbf{5}}$, which can thus be interpreted as that where the $\mathcal{B}-\mathcal{L}$ symmetry is explicitly broken.}
\label{Fig3}%
\end{figure}

With the $\mathbf{H}_{\mathbf{10}}$, the situation is a bit more
complicated. Naively, we would like to replace $\mathbf{H}_{\mathbf{50}}$ in
Eq.~(\ref{Majo50}) by $(\mathbf{H}_{\mathbf{10}}^{\dagger})^{2}$, which contains a
component transforming as $\mathbf{50}$. But since $\mathbf{H}_{\mathbf{10}}$
does not couple to fermions at the renormalizable level, we cannot couple
$\chi_{\mathbf{10}}\chi_{\mathbf{10}}\rightarrow\mathbf{H}_{\mathbf{10}%
}\mathbf{H}_{\mathbf{10}}$ directly. At an intermediate stage, a combination
of fields coupled to $\chi_{\mathbf{10}}$ and transforming as a $\mathbf{50} $
is needed. By inspection, the simplest such combination is $\varepsilon
_{VWABZ}h_{\mathbf{5}}^{Z}\mathbf{A}_{C}^{B}\mathbf{A}_{F}^{A}$, which can
then be coupled to $\mathbf{H}_{\mathbf{10}}^{2}$ thanks to the cubic
$\varepsilon_{ABCDE}\mathbf{H}_{\mathbf{10}}^{AB}\mathbf{H}_{\mathbf{10}}%
^{CD}h_{\mathbf{5}}^{E}$ interaction. Altogether, the two loop process
$\chi_{\mathbf{10}}\chi_{\mathbf{10}}\rightarrow\mathbf{A}_{\mathbf{24}%
}\mathbf{A}_{\mathbf{24}}h_{\mathbf{5}}^{\dagger}\rightarrow\mathbf{H}_{\mathbf{10}%
}\mathbf{H}_{\mathbf{10}}$ can thus proceed, see Fig.~\ref{Fig3}, and generates the effective action
term%
\begin{equation}
\mathcal{L}_{\dim5}^{eff}=\frac{C_{\dim5}}{v_{10}}(\bar{\chi}_{\mathbf{10}%
}^{c})^{AB}(\mathbf{H}_{\mathbf{10}}^{\dagger})_{AB}(\mathbf{H}_{\mathbf{10}%
}^{\dagger})_{CD}(\chi_{\mathbf{10}})^{CD}\ ,\ \ C_{\dim5}\sim\frac{\gamma
\mathbf{Y}_{10}g^{4}}{(4\pi)^{4}v_{10}}\ ,\label{MajoH10}%
\end{equation}
with the scale set by the GUT symmetry breaking. Once the $\mathbf{H}%
_{\mathbf{10}}$ acquires its expectation value, we arrive at a Majorana mass
term for the right-handed neutrinos%
\begin{equation}
\mathcal{L}_{10}^{eff}\sim\mathcal{O}(1)\times\frac{g^{4}}{(4\pi)^{4}}%
v_{10}\times\bar{\nu}_{R}^{c}\nu_{R} \ .
\label{MajoH10b}
\end{equation}
This is the flipped $SU(5)$ version~\cite{Rodriguez:2013rma} of the Witten $SO(10)$ mechanism~\cite{Witten:1979nr}. Overall,
the right-handed neutrinos are significantly lighter than in with
$\mathbf{H}_{\mathbf{50}}$, and $\mathbf{Y}_{10}$ need not be suppressed to
account for the measured neutrino mass splittings.

\subsection{PQ-flipped axion model}

There is one particular aspect of the symmetries of the flipped model that is
somewhat obscured in the above description, but that will play an important
role now. In the absence of the $\gamma$ coupling in the potential of
Eq.~(\ref{PotH10H10h5}), the theory actually has a larger symmetry. It is
invariant under two separate $U(1)$, since nothing relates the $U(1)_{X}%
\rightarrow U(1)^{\mathbf{10}}$ under which $\mathbf{H}_{\mathbf{10}}$ is
charged to the $U(1)_{X}\rightarrow U(1)^{\mathbf{5}}$ exhibited by the Yukawa
couplings of Eq.~(\ref{YukFlip}). The gauged symmetry is then embedded in the
product, $U(1)_{X}\subset U(1)^{\mathbf{10}}\otimes U(1)^{\mathbf{5}}$,
exactly as the gauged hypercharge is embedded in the $U(1)_{1}\otimes
U(1)_{2}$ symmetry in the THDM, see Eq.~(\ref{PatternTHDM}).

Though the symmetry is enlarged when $\gamma=0$, there is no extra Goldstone
boson because without $\gamma$, $\mathcal{B}-\mathcal{L}$ emerges as an exact
global symmetry at the low scale.\ Indeed, since $\mathbf{H}_{\mathbf{10}}$ is
neutral under $U(1)^{\mathbf{5}}$, that part of the symmetry stays unbroken
until the electroweak scale, and $U(1)_{\mathcal{B}-\mathcal{L}}$ arises from
$U(1)^{\mathbf{5}}\otimes U(1)_{Y}$ after the fiveplet acquires its VEV,
exactly as in the minimal $SU(5)$ model. Note, by the way, that the $\nu
_{R}\leftrightarrow e_{R}$ and $u_{R}\leftrightarrow d_{R}$ interchanges are
$\mathcal{B}-\mathcal{L}$ invariant, so the dimension-six leptoquark
interactions of the flipped model still conserve $\mathcal{B}-\mathcal{L}$.
Depending on $\gamma$, the patterns of the $U(1)$ symmetry breaking in the
flipped models are thus
\begin{equation}
\begin{array}
[c]{rcc}%
\gamma=0: & U(1)^{\mathbf{10}}\otimes U(1)^{\mathbf{5}}\otimes U(1)_{Y}%
\rightarrow U(1)_{\mathcal{B}-\mathcal{L}} & :2~\text{Goldstone bosons\ ,}\\
\gamma\neq0: & U(1)_{X}\otimes U(1)_{Y}\rightarrow\varnothing &
:2~\text{Goldstone bosons\ .}%
\end{array}
\end{equation}
Note that strictly speaking, $U(1)_{Y}$ is not entirely broken since the
unbroken $U(1)_{em}$ emerges from a combination of $U(1)_{Y}$ and $U(1)\subset
SU(2)_{L}$, but this is not essential. In both cases, the two Goldstone bosons
are the WBG of the $X^{0}$ and $Z^{0}$. So, one should really understand the
$\gamma$ coupling as the source of lepton-number violation. When $\gamma\neq
0$, by enforcing $U(1)^{\mathbf{10}}\otimes U(1)^{\mathbf{5}}\rightarrow
U(1)_{X}$, the fermionic symmetry from which the $\mathcal{B}-\mathcal{L}$
symmetry would emerge becomes that broken by $\mathbf{H}_{\mathbf{10}}$, and
thus never arises. That is the key to induce the $\Delta\mathcal{L}=2$
Majorana mass term dynamically. The same reasoning can be made with the
$\mathbf{H}_{\mathbf{50}}$ instead of the $\mathbf{H}_{\mathbf{10}}$. One must
introduce either the $\mathbf{H}_{\mathbf{50}}\mathbf{H}_{\mathbf{50}%
}^{\dagger}\mathbf{H}_{\mathbf{50}}h_{\mathbf{5}}^{\dagger}$ quartic coupling
or directly the Majorana coupling $\bar{\chi}_{\mathbf{10}}^{c%
}Y_{\mathbf{50}}\chi_{\mathbf{10}}\mathbf{H}_{\mathbf{50}}$ to force
$U(1)^{\mathbf{50}}\otimes U(1)^{\mathbf{5}}\rightarrow U(1)_{X}$, otherwise
the $\mathcal{B}-\mathcal{L}$ symmetry again arises at the low scale and the
seesaw mechanism is forbidden.

With the above symmetries in mind, to introduce the axion, we proceed as in
the PQ model and add a second Higgs fiveplet:%
\begin{align}
\mathcal{L}_{\text{Yukawa}}  & =-\frac{1}{4}\varepsilon_{ABCDE}(\bar{\chi
}_{\mathbf{10}}^{c})^{AB}\mathbf{Y}_{10}(\chi_{\mathbf{10}}%
)^{CD}h_{2,\mathbf{5}}^{E}\\
& +\sqrt{2}(\bar{\psi}_{\mathbf{\bar{5}}}^{c})_{A}\mathbf{Y}_{5}%
(\chi_{\mathbf{10}})^{AB}(h_{1,\mathbf{5}}^{\dagger})_{B}+(\bar{\psi
}_{\mathbf{\bar{5}}}^{c})_{A}\mathbf{Y}_{1}^{T}\psi_{\mathbf{1}%
}(h_{2,\mathbf{5}})^{A}+h.c.\;.
\end{align}
There is some freedom in deciding which fiveplet contributes to each coupling.
The above choice, which differs from that of the minimal model in
Eq.~(\ref{SU5THDM}), permits to recover the THDM of type II after the $SU(5)$
breaking:%
\begin{equation}
\mathcal{L}_{\text{Yukawa}}\rightarrow-\bar{u}_{R}\mathbf{Y}_{5}q_{L}\Phi
_{1}-\bar{d}_{R}\mathbf{Y}_{10}q_{L}\Phi_{2}^{\dagger}-\bar{e}_{R}%
\mathbf{Y}_{1}\ell_{L}\Phi_{2}^{\dagger}-\nu_{R}\mathbf{Y}_{5}^{T}\ell_{L}%
\Phi_{1}+h.c.\ .\label{YukFlippedTHDM}%
\end{equation}
With appropriate charges, these Yukawa couplings are invariant under the
independent rephasing of the Higgs fiveplets. Thus, at this level, the
$U(1)^{\mathbf{5}}$ symmetry is enlarged into $U(1)_{1}^{\mathbf{5}}\otimes
U(1)_{2}^{\mathbf{5}}$.

Let us assume the scalar potential shares this symmetry, except for the two
possible $\gamma$ couplings:%
\begin{equation}
V\supset\varepsilon_{ABCDE}\mathbf{H}_{\mathbf{10}}^{AB}\mathbf{H}%
_{\mathbf{10}}^{CD}(\gamma_{1}h_{1,\mathbf{5}}^{E}+\gamma_{2}h_{2,\mathbf{5}%
}^{E})+h.c.\ .\label{GammaCoup}%
\end{equation}
Depending on which of these couplings is non-zero, various symmetry-breaking
patterns are realized:
\[%
\begin{array}
[c]{rcc}%
\gamma_{1}=\gamma_{2}=0: & U(1)^{\mathbf{10}}\otimes U(1)_{1}^{\mathbf{5}%
}\otimes U(1)_{2}^{\mathbf{5}}\otimes U(1)_{Y}\rightarrow U(1)_{\mathcal{B}%
-\mathcal{L}} & :3~\text{Goldstone bosons\ ,}\\
\gamma_{1}\neq0\text{ or }\gamma_{2}\neq0: & U(1)_{1}^{\mathbf{5}}\otimes
U(1)_{2}^{\mathbf{5}}\otimes U(1)_{Y}\rightarrow\varnothing &
:3~\text{Goldstone bosons\ ,}\\
\gamma_{1}\neq0\text{ and }\gamma_{2}\neq0: & U(1)_{X}\otimes U(1)_{Y}%
\rightarrow\varnothing & :2~\text{Goldstone bosons\ .}%
\end{array}
\]
For the first two scenarios, though the number of Goldstone bosons is the same
as in the usual DFSZ model, the situation is very different. Indeed, first,
all three electrically neutral pseudoscalar degrees of freedom $\eta_{10}$,
$\eta_{1,5}$, $\eta_{2,5}$ are massless, and second, two combinations have to
be used to form the WBG of the $X^{0}$ and $Z^{0}$ gauge bosons:%
\begin{equation}
G_{Y}^{0}\sim v_{1}\eta_{1,5}+v_{2}\eta_{2,5}\ ,\ \ G_{X}^{0}\sim-2v_{1}%
\eta_{1,5}-2v_{2}\eta_{2,5}+v_{10}\eta_{10}\ .\label{GXGY}%
\end{equation}
Those states are not immediately orthogonal because of the mixing induced by
$Y=-Y^{\prime}/5+2X/5$, but this does not matter. The point is that the
remaining Goldstone boson must be orthogonal to both these WBG, and the only
orthogonal direction available is
\begin{equation}
a^{0}\sim\frac{\eta_{1,5}}{v_{1}}-\frac{\eta_{2,5}}{v_{2}}%
\ .\label{FlipEWaxion}%
\end{equation}
Thus, the axion cannot have a $\eta_{10}$ component, and its dynamics entirely
take place at the electroweak scale. This is confirmed looking at the third
scenario, with $\gamma_{1}\neq0$ and $\gamma_{2}\neq0$.\ In that case, both
fiveplets have the same $X$ and $Y$ charges, and thus nothing prevents a
$h_{1,\mathbf{5}}h_{2,\mathbf{5}}^{\dagger}$ coupling in the potential. Even
if it is not initially present, it is radiatively induced by the $\gamma_{1}$
and $\gamma_{2}$ couplings. But to such a $h_{1,\mathbf{5}}h_{2,\mathbf{5}%
}^{\dagger}$ coupling corresponds the pseudoscalar potential term (in the
polar representation)%
\begin{equation}
V(\eta_{1},\eta_{2},\eta_{10})\sim\cos\left(  \frac{\eta_{1,5}}{v_{1}}%
-\frac{\eta_{2,5}}{v_{2}}\right)  \ .
\end{equation}
Without surprise, the state becoming massive is the axion since the WBG of the
$X^{0}$ and $Z^{0}$ gauge bosons cannot do so.

The symmetry breaking patterns are confirmed by working out the charges
explicitly. For the first scenario, with $\gamma_{1}=\gamma_{2}=0$, the Yukawa
couplings impose%
\begin{equation}%
\begin{tabular}
[c]{ccccccc}\hline
$\gamma_{1}=\gamma_{2}=0$ & $\mathbf{H}_{\mathbf{10}}$ & $h_{1,\mathbf{5}}$ &
$h_{2,\mathbf{5}}$ & $\chi_{\mathbf{10}}$ & $\psi_{\mathbf{\bar{5}}}$ &
$\psi_{\mathbf{1}}$\\\hline
$U(1)^{\mathbf{10}}$ & $1$ & $0$ & $0$ & $0$ & $0$ & $0$\\
$U(1)_{1}^{\mathbf{5}}$ & $0$ & $1$ & $0$ & $0$ & $1$ & $-1$\\
$U(1)_{2}^{\mathbf{5}}$ & $0$ & $0$ & $1$ & $-1/2$ & $1/2$ & $-3/2$\\\hline
\end{tabular}
\end{equation}
The PQ charge of the doublets $\Phi_{1}\subset h_{1,\mathbf{5}}^{\dagger}$
and $\Phi_{2}\subset h_{2,\mathbf{5}}^{\dagger}$ are the same as before (note
the $\dagger$ though), $x$ and $-1/x$, and so are the PQ charges derived from
Eq.~(\ref{YukFlippedTHDM}) which match Eq.~(\ref{PQferm}). This means that
from the GUT-scale $U(1)$s, we can construct%
\begin{align}
X &  =U_{1}^{\mathbf{10}}-2U_{1}^{\mathbf{5}}-2U_{2}^{\mathbf{5}}\ ,\\
\mathcal{B}-\mathcal{L} &  =-2U_{1}^{\mathbf{5}}-2U_{2}^{\mathbf{5}}-2Y\ ,\\
PQ &  =\zeta_{10}^{PQ}U_{1}^{\mathbf{10}}+\zeta_{1}^{PQ}U_{1}^{\mathbf{5}%
}+\zeta_{2}^{PQ}U_{2}^{\mathbf{5}}+\zeta_{Y}^{PQ}Y\ ,
\end{align}
with%
\begin{equation}
\zeta_{10}^{PQ}=0\ ,\ \ \zeta_{1}^{PQ}=2\beta+x-\frac{1}{x}\ ,\ \ \zeta
_{2}^{PQ}=2\beta+2x\ ,\ \ \zeta_{Y}^{PQ}=2\beta+2x-\frac{1}{x}\ .
\label{PQflippedSol}
\end{equation}
The PQ charge of the fermions are the same as in the minimal model,
Eq.~(\ref{PQFinalCharge}), with in addition $PQ(\nu_{R})=\beta+x$. This
reflects the fact that the Yukawa couplings remain those of a THDM of type II.
Also, they exhibit the same one-parameter ambiguity originating in the global
$\mathcal{B}-\mathcal{L}$ invariance. Note that when this symmetry is active, and only in that case, it is possible to choose $SU(5)$ invariant PQ charges by setting $\zeta_{Y}^{PQ} = 0$, i.e., 
\begin{equation}
\beta = -x+\frac{1}{2x}\ .
\label{PQflipSU5}
\end{equation}
If only one $\gamma$ is non-zero, the patterns of charges are%
\begin{equation}%
\begin{tabular}
[c]{ccccccc}\hline
$\gamma_{1}=0$ or $\gamma_{2}=0$ & $\mathbf{H}_{\mathbf{10}}$ &
$h_{1,\mathbf{5}}$ & $h_{2,\mathbf{5}}$ & $\chi_{\mathbf{10}}$ &
$\psi_{\mathbf{\bar{5}}}$ & $\psi_{\mathbf{1}}$\\\hline
$U(1)_{1}^{\mathbf{5}}$ & $-1/2\delta_{\gamma_{2}}^{0}$ & $1$ & $0$ & $0$ &
$1$ & $-1$\\
$U(1)_{2}^{\mathbf{5}}$ & $-1/2\delta_{\gamma_{1}}^{0}$ & $0$ & $1$ & $-1/2 $
& $1/2$ & $-3/2$\\\hline
\end{tabular}
\end{equation}
Though $\mathbf{H}_{\mathbf{10}}$ is charged under $U(1)_{1}^{\mathbf{5}%
}\otimes U(1)_{2}^{\mathbf{5}}$, the PQ charge of its electrically-neutral
component vanishes since the axion has no $\eta_{10}$ component, while the PQ
charges of $\Phi_{1}\subset h_{1,\mathbf{5}}^{\dagger}$ and $\Phi_{2}\subset
h_{2,\mathbf{5}}^{\dagger}$ are the same as before. These charges can be
rearranged into
\begin{align}
X &  =-2U_{1}^{\mathbf{5}}-2U_{2}^{\mathbf{5}}\ ,\\
\mathcal{B}-\mathcal{L} &  =-2U_{1}^{\mathbf{5}}-2U_{2}^{\mathbf{5}}-2Y\ ,\\
PQ &  =\zeta_{1}^{PQ}U_{1}^{\mathbf{5}}+\zeta_{2}^{PQ}U_{2}^{\mathbf{5}}%
+\zeta_{Y}^{PQ}Y\ .
\end{align}
For $\mathcal{B}-\mathcal{L}$, the above combination reproduces the charges of the fermions, but not that of the scalars since $\mathbf{H}_{\mathbf{10}}$ ends up with  $\mathcal{B}-\mathcal{L} = 1$. This shows explicitly that this symmetry is broken by the $\gamma$ couplings. The coefficients $\zeta_{i}^{PQ}$ are now uniquely defined:%
\begin{equation}%
\begin{array}
[c]{llll}%
\gamma_{1}=0: & \zeta_{1}^{PQ}=-x-\dfrac{1}{x}\ , & \zeta_{2}^{PQ}=0\ ,\  &
\zeta_{Y}^{PQ}=-\dfrac{1}{x}\ ,\\
\gamma_{2}=0: & \zeta_{1}^{PQ}=0\ ,\ \  & \zeta_{2}^{PQ}=x+\dfrac{1}{x}\ , &
\zeta_{Y}^{PQ}=x\ .
\end{array}
\end{equation}
Consequently, the PQ charges of the fermions are unambiguously defined, and
correspond to Eq.~(\ref{PQFinalCharge}) with either
\begin{equation}%
\gamma_{1}=0: \beta=-x \  , \ \ \ \gamma_{2}=0: \beta=-\frac{1}{2} \left(  x-\frac{1}{x}\right)   \ .
\label{betagammaPQ}
\end{equation}

Concerning neutrinos, since $PQ(\nu_{R})=\beta+x$, we find $PQ(\nu_{R})=0$ when $\gamma_{1}=0$, and $PQ(\nu_{R})=1/2(x+1/x)$ when $\gamma_{2}=0$. This means that the mechanism in Eq.~(\ref{MajoH10}) to induce a Majorana neutrino mass term works only when $\gamma_{2}\neq0$. This is not surprising looking at Fig.~\ref{Fig3}: the cubic scalar coupling has to involve the same fiveplet as that coupled to $\bar{\chi}_{\mathbf{10}}^{c}\chi_{\mathbf{10}}$. If that is not the case, a Majorana mass term can still be induced since $\mathcal{B}-\mathcal{L}$ is broken. But with $\gamma_{2}=0$, one has to rely on more complicated processes involving the quartic $\alpha(h_{1,\mathbf{5}}^{\dagger}h_{1,\mathbf{5}}) (h_{2,\mathbf{5}}^{\dagger}h_{2,\mathbf{5}})$. Instead of Eq.~(\ref{MajoH10}), a Majorana mass term then arises at the dimension-seven level%
\begin{equation}
\mathcal{L}_{\dim7}^{eff}=\frac{C_{\dim7}}{v_{10}^{3}}\bar{\chi}_{\mathbf{10}%
}^{c}(\mathbf{H}_{\mathbf{10}}^{\dagger})^{2}h_{1,\mathbf{5}}^{\dagger
}h_{2,\mathbf{5}}\chi_{\mathbf{10}}+h.c.\ .
\label{MajoDim7}
\end{equation}
With the Majorana mass scale $M_{R}\sim v_{5}^{2}/v_{10}$, this scenario is
not physically viable because $M_{R}$ is not large enough to compensate the
neutrino Dirac mass term in Eq.~(\ref{YukFlippedTHDM}).

\subsection{DFSZ-flipped axion models}

The flipped model needs a further GUT-scale scalars to move the axion up from
the electroweak scale. The simplest solutions, in terms of additional fields,
are either a singlet or another 10. Let us analyze these two scenarios in turn.

\subsubsection{Singlet DFSZ}

With a singlet, several new couplings can occur in the scalar potential. To
investigate the possible realizations of the DFSZ mechanism, and besides the
couplings immediately invariant under the separate rephasing of the fields,
let us add the following mixing terms%
\begin{equation}
\mathcal{L}_{DFSZ}\supset\lambda\phi_{\mathbf{1}}^{2}h_{1,\mathbf{5}}%
^{\dagger}h_{2,\mathbf{5}}+\varepsilon_{ABCDE}\mathbf{H}_{\mathbf{10}}%
^{AB}\mathbf{H}_{\mathbf{10}}^{CD}(\gamma_{1}^{\phi}h_{1,\mathbf{5}}^{E}%
\phi_{\mathbf{1}}^{\dagger}+\gamma_{2}^{\phi}h_{2,\mathbf{5}}^{E}%
\phi_{\mathbf{1}})+h.c.\ .
\label{FlipSingDFS}
\end{equation}
Also, we assume that the dimensionless $\gamma_{i}^{\phi}$ couplings replace
those in Eq.~(\ref{GammaCoup}). The reasons for choosing these particular couplings will be apparent shortly. 

There are four patterns of $U(1)$ symmetry breaking, depending on which couplings are present. Let us discuss each case in turn.
\begin{itemize}
\item  $\mathbf{\lambda = \gamma_{1}^{\phi} =\gamma_{2}^{\phi}=0 }$: There are many active $U(1)$s in this case, 
\begin{equation}
U(1)^{\mathbf{1}}\otimes U(1)^{\mathbf{10}}\otimes U(1)_{1}^{\mathbf{5}}\otimes U(1)_{2}^{\mathbf{5}%
}\otimes U(1)_{Y}\rightarrow U(1)_{\mathcal{B}-\mathcal{L}} \ ,
\label{FlipSinglet}
\end{equation}
and four Goldstone bosons. Two of them are eaten by the $X^{0}$ and $Z^{0}$ gauge bosons. The one embedded in the singlet decouples entirely since there is nothing relating $\phi_{\mathbf{1}}$ to the rest of the dynamics. The axion remains at the electroweak scale, see Eq.~(\ref{FlipEWaxion}), and no Majorana mass term is allowed. The PQ charges are the same as in the PQ model of the previous section, with the parameter $\beta$ free.

\item $\mathbf{\lambda=0}$\textbf{ and either }$\mathbf{\gamma_{1}^{\phi}=0}$\textbf{ or }$\mathbf{\gamma_{2}^{\phi}=0}$: The symmetry breaking chain is now
\begin{equation}
U(1)^{\mathbf{10}}\otimes U(1)_{1}^{\mathbf{5}}\otimes U(1)_{2}^{\mathbf{5}%
}\otimes U(1)_{Y}\rightarrow\varnothing \ .
\end{equation}
Because the $\gamma_{i}^{\phi}$ couplings do not contribute to pseudoscalar masses, the Goldstone bosons are the same as in the previous case, and the axion remains the electroweak-scale state of Eq.~(\ref{FlipEWaxion}). The only difference is that the pseudoscalar state in $\phi_{\mathbf{1}}$ can now be identified as a pure, PQ-neutral majoron. Indeed, with the presence of $\phi_{\mathbf{1}}$ in the $\gamma_{1}^{\phi}$ or $\gamma_{2}^{\phi}$ coupling, the equivalent of Fig.~\ref{Fig3} leads to a $\nu_{R}$ Majorana mass term only once $\phi_{\mathbf{1}}$ acquires its VEV. Overall, the PQ charges are the same as in the PQ model since $PQ(\phi_{\mathbf{1}})=0$, and depending on which $\gamma_{i}^{\phi}$ is non-zero, $\beta$ is fixed as in Eq.~(\ref{betagammaPQ}).

\item $\mathbf{\lambda \neq 0}$\textbf{ with }$\mathbf{\gamma_{1}^{\phi}=\gamma_{2}^{\phi}=0}$: Since the $\gamma_{i}^{\phi}$ couplings are absent, $\mathcal{B}-\mathcal{L}$ is active and the symmetry-breaking chain is
\begin{equation}
U(1)^{\mathbf{10}}\otimes U(1)_{1}^{\mathbf{5}}\otimes U(1)_{2}^{\mathbf{5}}\otimes
U(1)_{Y}\rightarrow U(1)_{\mathcal{B}-\mathcal{L}} \ .
\end{equation}
There are thus three Goldstone bosons. The role of the $\lambda$ coupling is to make one pseudoscalar state massive
\begin{equation}
\pi^{0}\sim -\frac{\eta_{1,5}}{v_{1}}+\frac{\eta_{2,5}}{v_{2}}+2\frac{\eta_{s}%
}{v_{s}}\ ,
\end{equation}
where $v_{s}$ is the VEV of the singlet $\phi_{\mathbf{1}}$, and $\eta_{s}$ its pseudoscalar component. Then, the only state orthogonal to $\pi^{0}$, $G_{X}^{0}$ and $G_{Y}^{0}$ (given in Eq.~(\ref{GXGY})) is necessarily%
\begin{equation}
a^{0}\sim\eta_{s}+\frac{v\sin2\beta}{v_{s}}(\cos\beta\eta_{1,5}-\sin\beta
\eta_{2,5})\ ,\label{FlipDFSZaxion}%
\end{equation}
where $v_{1}=v_{5}\sin\beta$, $v_{2}=v_{5}\cos\beta$, exactly as in Eq.~(\ref{DFSZaxion}). The PQ charges of the scalars are then uniquely defined, with $PQ(\Phi_{1},\Phi_{2})=(x,-1/x)$, while those of the fermions are again given by Eq.~(\ref{PQFinalCharge}), including the $\beta$ ambiguity since $\mathcal{B}-\mathcal{L}$ is active. 

\item \textbf{At least two non-vanishing couplings among }$\mathbf{\lambda}$\textbf{ and }$\mathbf{\gamma_{1,2}^{\phi}}$: All these cases are equivalent because whichever two couplings are present, the third can be induced radiatively. Thus, the symmetry breaking pattern in the same in all these cases:
\begin{equation}
U(1)_{1}^{\mathbf{5}}\otimes U(1)_{2}^{\mathbf{5}}\otimes U(1)_{Y}\rightarrow\varnothing \ .
\label{flipnDFSZ}
\end{equation}
The Goldstone bosons are not affected by the $\mathbf{\gamma_{1,2}^{\phi}}$ couplings, the axion remains as in Eq.~(\ref{FlipDFSZaxion}), and the PQ charges of all the scalars but $\mathbf{H}_{\mathbf{10}}$ are unchanged. For fermions, however, the PQ charge ambiguity is removed and $\beta = -1/4(3x-1/x)$. This value is precisely that which ensures the Majorana mass operator arising from the analogue of Fig.~\ref{Fig3} to be allowed by the PQ symmetry: 
\begin{equation}
\mathcal{L}_{\dim6}^{eff}=\frac{C_{\dim6}}{v_{10}^{2}}\bar{\chi}_{\mathbf{10}%
}^{c}(\mathbf{H}_{\mathbf{10}}^{\dagger})^{2}\phi_{\mathbf{1}}^{\dagger
}\chi_{\mathbf{10}}+h.c.\ .
\end{equation}
Phenomenologically, when $v_s \approx v_{10}$, this dimension-six operator is equivalent to the dimension-five one of Eq.~(\ref{MajoH10}). Overall, this scenario can be viewed as the flipped analog of the $\nu$DFSZ of the minimal $SU(5)$ model, see Eq.~(\ref{LagrnuDFSZ}). Here also, the axion is identified with the Majoron. 
\end{itemize}

Note, finally, that we can now understand the specific choice of couplings made in Eq.~(\ref{FlipSingDFS}). For the axion to remain massless, they have to be compatible among themselves. For example, if one adds to $\gamma_{1}^{\phi}$ and $\gamma_{2}^{\phi}$ the coupling $\phi_{\mathbf{1}}^{2}h_{1,\mathbf{5}}h_{2,\mathbf{5}}^{\dagger}$ instead of $\phi_{\mathbf{1}}^{2}h_{1,\mathbf{5}}^{\dagger}h_{2,\mathbf{5}}$, then the symmetry pattern collapses to $U(1)_{X}\otimes U(1)_{Y}\rightarrow\varnothing$, with both $\pi^{0}$ and $a^{0}$ massive. Thus, the set in  Eq.~(\ref{FlipSingDFS}) is one example that ensures viable scenarios do exist. Though there are other possible sets of couplings that could produce an acceptable axion state, the symmetry breaking patterns would be very similar, and only the value of $\beta$ could be different.

\subsubsection{Fundamental DFSZ}

If instead of a singlet one introduces a second $H_{\mathbf{10}}$, with the same gauge quantum numbers, the possible mixing terms in the scalar potentials are%
\begin{equation}
\mathcal{L}_{DFSZ}\supset\gamma^{ijk}\mathbf{H}%
_{i,\mathbf{10}}\mathbf{H}_{j,\mathbf{10}}h_{k,\mathbf{5}}
+\alpha^{ijkl}(h_{i,\mathbf{5}}^{\dagger}h_{j,\mathbf{5}})\langle
\mathbf{H}_{k,\mathbf{10}}^{\dagger}\mathbf{H}_{l,\mathbf{10}}\rangle
+\beta^{ijkl}h_{i,\mathbf{5}}^{\dagger}\mathbf{H}_{k,\mathbf{10}}^{\dagger
}\mathbf{H}_{l,\mathbf{10}}h_{j,\mathbf{5}}\ ,
\label{EntangFund}
\end{equation}
where $i,j,k,l=1,2$. Whenever $i\neq j$ and/or $k\neq l$, the $U(1)$ charges of the scalar states get entangled. From this point of view, the $\alpha^{ijkl}$ and $\beta^{ijkl}$ couplings have exactly the same effect, so it is sufficient to consider only the former and we set $\beta^{ijkl}=0$. 

To set the stage, consider the situation without any of these couplings. The scalar potential is invariant under $U(1)_{1}^{\mathbf{10}}\otimes U(1)_{2}^{\mathbf{10}}\otimes U(1)_{1}^{\mathbf{5}}\otimes U(1)_{2}^{\mathbf{5}}$, and the $U(1)_{\mathcal{B}%
-\mathcal{L}}$ symmetry emerges at low energy. There are thus two extra Goldstone bosons besides those eaten by the gauge bosons, which are now%
\begin{equation}
G_{Y}^{0}\sim v_{1}\eta_{1,5}+v_{2}\eta_{2,5}\ ,\ \ G_{X}^{0}\sim-2v_{1}%
\eta_{1,5}-2v_{2}\eta_{2,5}+v_{1,10}\eta_{1,10}+v_{2,10}\eta_{2,10}\ .
\end{equation}
In this case, the axion $a_{EW}^{0}$ is given by Eq.~(\ref{FlipEWaxion}), and is accompanied by another massless states $a_{GUT}^{0}$, decoupled from the SM fermions,
\begin{equation}
a_{EW}^{0}\sim\frac{\eta_{1,5}}{v_{1}}-\frac{\eta_{2,5}}{v_{2}}\ ,\ \ a_{GUT}%
^{0}\sim\frac{\eta_{1,10}}{v_{1,10}}-\frac{\eta_{2,10}}{v_{2,10}%
}\ .\label{aEWGUT}%
\end{equation}
Neither of these states can be a viable axion candidate, so some of the mixing terms have to be turned on.

Consider first the quartic terms in the potential, keeping the $\gamma^{ijk}$ to zero. The trick to implement the DFSZ mechanism is to turn on just enough of these couplings to make a linear combination of $a_{EW}^{0}$ and $a_{GUT}^{0}$ massive.\ By orthogonality, the axion will then have a component in the $\eta_{i,10}$ direction, and this will move it up to the GUT scale. Clearly, turning on any of the $\alpha^{ijkl}$ with $i=j$ or $k=l$ does not work because the state becoming massive is either $a_{EW}^{0}$ or $a_{GUT}^{0}$, but not a combination of them. Instead, turning on $\alpha^{1212}$ or $\alpha^{2121}$ produces the massive%
\begin{equation}
\pi^{0}\sim\frac{\eta_{1,5}}{v_{1}}-\frac{\eta_{2,5}}{v_{2}}+\frac{\eta
_{1,10}}{v_{1,10}}-\frac{\eta_{2,10}}{v_{2,10}}\ ,
\end{equation}
and by orthogonality, the massless axion is then%
\begin{equation}
a^{0}\sim(-\cos\alpha \ \eta_{1,10}+\sin\alpha\ \eta_{2,10})\sin2\alpha+\frac
{v_{5}\sin2\beta}{v_{10}}(\cos\beta\ \eta_{1,5}-\sin\beta\ \eta_{2,5})\ ,
\end{equation}
where $\tan\alpha=v_{1,10}/v_{2,10}$ and $\tan\beta=v_{1}/v_{2}$. Comparing with Eq.~(\ref{DFSZaxion}) or~(\ref{FlipDFSZaxion}), this is precisely what we were after. From this expression, the corresponding PQ charges are immediately found to be, upon a proper normalization,%
\begin{equation}
PQ(\Phi_{1}\subset h_{1,\mathbf{5}}^{\dagger},\Phi_{2}\subset h_{2,\mathbf{5}%
}^{\dagger},\phi_{1}\subset\mathbf{H}_{1,\mathbf{10}},\phi_{2}%
\subset\mathbf{H}_{2,\mathbf{10}})=\left(  x,-\frac{1}{x},\left(
x+\frac{1}{x}\right)  \cos^{2}\alpha,-\left(  x+\frac{1}{x}\right)  \sin
^{2}\alpha\right)  \ .
\label{FlipFund}
\end{equation}
Therefore, the PQ charges of the fermions are still given by Eq.~(\ref{PQFinalCharge}), including the $\beta$ ambiguity since $\mathcal{B}-\mathcal{L}$ remains active when $\gamma^{ijk}=0$. The situation is similar turning on $\alpha^{1221}$ or $\alpha^{2112}$ instead, with only the PQ charges of $\phi_{i}\subset\mathbf{H}_{\mathbf{10}}^{i}$ changing sign. However, if say $\alpha^{1212}$ and $\alpha^{1221}$ are simultaneously present, then the $a_{EW}^{0}$ and $a_{GUT}^{0}$ states of Eq.~(\ref{aEWGUT}) both become massive.

Concerning the $\gamma^{ijk}$ couplings, note that none of them can directly generate a mass term for the pseudoscalars. Thus, if a single $\gamma^{ijk}$ coupling is present, but $\alpha^{ijkl}=0$,
there are again the two extra Goldstone bosons of Eq.~(\ref{aEWGUT}). Indeed, the initial symmetry has one less $U(1)$ because of the $\gamma^{ijk}$ coupling, but $U(1)_{\mathcal{B}-\mathcal{L}}$ no longer emerges at low
energy. In this case, $a_{GUT}^{0}$ is a pure majoron state, and the axion is stuck at the electroweak scale.

\begin{table}[t]
\centering
\begin{tabular}[c]{ccccc}\hline
& \multicolumn{2}{c}{$\alpha^{1212}$} & \multicolumn{2}{c}{$\alpha^{1221}$}\\
& $\beta$ & $(l,m,n,p)$ & $\beta$ & $(l,m,n,p)$\\\hline
\multicolumn{1}{l}{$\gamma^{111}$} & \multicolumn{1}{l}{$\left(  x+\dfrac
{1}{x}\right)  s_{\alpha}^{2}-\dfrac{3x}{2}-\dfrac{1}{2x}$} &
\multicolumn{1}{l}{$(1,0,1,0)$} & \multicolumn{1}{l}{$-\left(  x+\dfrac{1}%
{x}\right)  s_{\alpha}^{2}+\dfrac{x}{2}+\dfrac{3}{2x}$} &
\multicolumn{1}{l}{$(3,0,0,1)$}\\
\multicolumn{1}{l}{$\gamma^{112}$} & \multicolumn{1}{l}{$\left(  x+\dfrac
{1}{x}\right)  s_{\alpha}^{2}-2x-\dfrac{1}{x}$} &
\multicolumn{1}{l}{$(2,0,0,0)$} & \multicolumn{1}{l}{$-\left(  x+\dfrac{1}%
{x}\right)  s_{\alpha}^{2}+\dfrac{1}{x}$} & \multicolumn{1}{l}{$(2,0,0,0)$}\\
\multicolumn{1}{l}{$\gamma^{121}$} & \multicolumn{1}{l}{$\left(  x+\dfrac
{1}{x}\right)  s_{\alpha}^{2}-x$} & \multicolumn{1}{l}{$(0,0,2,0)$} &
\multicolumn{1}{l}{$-\left(  x+\dfrac{1}{x}\right)  s_{\alpha}^{2}+\dfrac
{1}{x}$} & \multicolumn{1}{l}{$(2,0,0,0)$}\\
\multicolumn{1}{l}{$\gamma^{122}$} & \multicolumn{1}{l}{$\left(  x+\dfrac
{1}{x}\right)  s_{\alpha}^{2}-\dfrac{3x}{2}-\dfrac{1}{2x}$} &
\multicolumn{1}{l}{$(1,0,1,0)$} & \multicolumn{1}{l}{$-\left(  x+\dfrac{1}%
{x}\right)  s_{\alpha}^{2}-\dfrac{x}{2}+\dfrac{1}{2x}$} &
\multicolumn{1}{l}{$(1,0,1,0)$}\\
\multicolumn{1}{l}{$\gamma^{221}$} & \multicolumn{1}{l}{$\left(  x+\dfrac
{1}{x}\right)  s_{\alpha}^{2}-\dfrac{x}{2}+\dfrac{1}{2x}$} &
\multicolumn{1}{l}{$(0,1,3,0)$} & \multicolumn{1}{l}{$-\left(  x+\dfrac{1}%
{x}\right)  s_{\alpha}^{2}-\dfrac{x}{2}+\dfrac{1}{2x}$} &
\multicolumn{1}{l}{$(1,0,1,0)$}\\
\multicolumn{1}{l}{$\gamma^{222}$} & \multicolumn{1}{l}{$\left(  x+\dfrac
{1}{x}\right)  s_{\alpha}^{2}-x$} & \multicolumn{1}{l}{$(0,0,2,0)$} &
\multicolumn{1}{l}{$-\left(  x+\dfrac{1}{x}\right)  s_{\alpha}^{2}-x$} &
\multicolumn{1}{l}{$(0,0,2,0)$}\\\hline
\end{tabular}
\label{FundDFSZ}
\caption{Values of $\beta$ for various viable implementations of the DFSZ mechanism using two $\mathbf{H}_{\mathbf{10}}$ multiplets, where $s_{\alpha}\equiv\sin \alpha$. In each case, the set of numbers refer to the exponents of the seesaw operator in Eq.~(\ref{FundSee}).}
\end{table}

The simplest viable scenarios where $\mathcal{B}-\mathcal{L}$ is broken use either one of the $\gamma^{ijk}$ together with one $\alpha^{lmnp}$ with $l\neq m$ and $n\neq p$, or two of the $\gamma^{ijk}$ couplings. These two situations are actually equivalent because a pair of  $\gamma^{ijk}$ couplings radiatively induces an effective $\alpha^{lmnp}$-like coupling, and a $\gamma^{ijk}$ coupling together with $\alpha^{lmnp}$ radiatively induces an effective $\gamma^{ijk}$-like coupling. By this we mean that though these effective couplings can be quite complicated, and may involve also the other quartic couplings of the scalar potential, they impose the same entanglement of the scalar charges as one of the $\gamma^{ijk}$ or $\alpha^{lmnp}$  coupling.

So, it is sufficient to cover all the cases to consider only one of the $\gamma^{ijk}$ couplings together with one of the $\alpha^{lmnp}$ couplings. The axion state and PQ charges are the same as with only the $\alpha^{lmnp}$ coupling, except that the ambiguity parameter $\beta$ is fixed, see Table~\ref{FundDFSZ}. Clearly, all these scenarios have the same axion phenomenology since they differ only in the parameter $\beta$, which now depends on both the GUT and EW-scale ratios of VEVs $v_{1,10}/v_{2,10}$ and $v_{1}/v_{2}$.

The same is true in the neutrino sector. We give in Table~\ref{FundDFSZ} the sets $(l,m,n,p)$ corresponding to the leading operator giving rise to the Majorana mass term for the right-handed neutrinos, i.e.,  
\begin{equation}
\mathcal{L}_{eff}=\frac{1}{v_{10}^{l+m+n+p-1}}\bar{\chi}_{\mathbf{10}}^{\mathrm{C}}(\mathbf{H}_{1,\mathbf{10}%
}^{\dagger})^{l}(\mathbf{H}_{1,\mathbf{10}})^{m}(\mathbf{H}_{2,\mathbf{10}%
}^{\dagger})^{m}(\mathbf{H}_{2,\mathbf{10}})^{p}\chi_{\mathbf{10}}\ ,
\label{FundSee}
\end{equation}
with $l-m+n-p = 2$. These operators are found from their invariance under the $U(1)_{1}^{\mathbf{5}}\otimes U(1)_{2}^{\mathbf{5}}$ symmetry. Contrary to the PQ model, see Eq.~(\ref{MajoDim7}), it is always possible to construct them using only the GUT-scale scalar decuplets, even when the $\gamma$ coupling does not involve the same fiveplet as $\mathbf{Y}_{10}$, because it is always possible to use the $\alpha$ coupling to switch from one fiveplet to the other (see Fig.~\ref{Fig3}). Once the decuplets acquire their VEVs, and except for extreme values of $v_{1,10}/v_{2,10}$, all these operators are clearly equivalent phenomenologically.

\section{Conclusions\label{Ccl}}

Axion models are based on the spontaneous breaking of an extra $U(1)$ symmetry. If this symmetry has a strong anomaly, the axion, the associated Goldstone boson, ends up coupled to gluons, and this ensures driving the strong CP violation to zero in the non-perturbative regime of QCD.

A characteristic feature of axion models is that the true $U(1)_{PQ}$ symmetry corresponding to the axion
is not trivial to identify, because of the presence of several other $U(1)$ symmetries acting on the same fields: baryon number $\mathcal{B}$, lepton number $\mathcal{L}$, and weak hypercharge. As a consequence, the
PQ charges can only be defined after $U(1)_{Y}$ is spontaneously broken, and even then, those of the fermions remain ambiguous whenever baryon or lepton number is conserved~\cite{Quevillon:2020hmx}. Specifically, given Yukawa couplings to two electroweak Higgs doublets of type II (see Eq.~(\ref{YukQuark})), the PQ charge of the SM fermions are expressed in function of the two free parameters, $\alpha$ and $\beta$, as:
\begin{equation}
PQ(q_{L},u_{R},d_{R},\ell_{L},e_{R})=\left( \alpha,\alpha+x,\alpha+\frac{1}{x},\beta,\beta+\frac{1}{x} \right)\ \ ,
\end{equation}
where $x =v_2/v_1$ and $v_i$ the VEV of each Higgs doublet. Our purpose was to study how these ambiguities manifest themselves in the $SU(5)$ GUT setting, see when they can be lifted, and how they permit to accommodate for a Majorana mass term for the neutrinos. Our main results are:

\begin{itemize}
\item In a GUT setting, one of the two ambiguities immediately disappears, and
\begin{equation}
3\alpha+\beta=-\left(  x+\frac{1}{x}\right)  \equiv2\mathcal{N}_{SU(5)}\ .%
\end{equation}
This can be understood either as a consequence of the $SU(5)$ gauge interactions breaking $\mathcal{B}+\mathcal{L}$, or because the anomalous couplings of the axion to all the SM gauge bosons must originate from the single anomaly coefficients of the global $SU(5)$ chiral currents, see Eq.~(\ref{unifano1}) and~(\ref{unifano2}). Remains thus only one freedom in the fermion PQ charges, $\beta$, corresponding to the conserved $\mathcal{B}-\mathcal{L}$ symmetry. In the Table~\ref{Tablesum} we summarise the status of this parameter for the models studied in details in this paper.

\item  The $\mathcal{B}$ and $\mathcal{L}$ are not exact symmetries at the GUT scale, but only emerge at the low scale. This means the PQ symmetry has to be defined similarly if it is to be compatible with $\mathcal{B}$ and/or $\mathcal{L}$ violating effects, as required for example to allow for a Majorana neutrino mass term. As a corollary, this means there is no reason to expect PQ charges to be invariant over $SU(5)$ multiplets, and indeed in most cases, they are not. It is only in the absence of neutrino masses, when $\mathcal{B}-\mathcal{L}$ is active, that $SU(5)$-invariant PQ charges can be defined, see Table~\ref{Tablesum}. As an aside, we also clarified the status of $\mathcal{B}$ and $\mathcal{L}$ in the flipped $SU(5)$ model, putting it on a par with the minimal model.

\item Table~\ref{Tablesum} shows that many different implementations were studied, but they all reproduce the same PQ charges, up to the value of $\beta$. This is true for both the minimal and flipped $SU(5)$ model, for various embedding of the axion in DFSZ-like models, and in the presence of a seesaw mechanism of type I, II, or when the DFSZ singlet also plays the role of the majoron. Though this fact can be understood as the orthogonality condition among Goldstone bosons stays essentially the same, and so are the low-energy Yukawa couplings, it is often obscured by the normalization of the PQ charges. Yet, this is remarkable because it means the low-energy phenomenology of the axion is the same in all these models, since as shown in Ref.~\cite{Quevillon:2019zrd}, its couplings to fermions and gauge bosons are independent of the value of $\beta$.
\end{itemize}

The strategy to construct viable embeddings of the axion within grand unified scenarios is thus clear. One must first identify precisely the GUT scale global and local symmetries. Then, the PQ symmetry, along with $\mathcal{B}$ and/or $\mathcal{L}$ if not explicitly broken, arise as a combination of the generators of all the $U(1)$ symmetries active at the GUT scale, including the weak hypercharge. This combination is fixed, up to some possible ambiguities, by the orthogonality requirement among the pseudoscalar states, including both the massive states and the would-be Goldstone bosons. As a final step, knowing the PQ charges of the scalars, one can derive those of the fermions. The main advantage of proceeding in this way, instead of first fixing the PQ charges of the scalars and fermions, is that automatically, enough room is left to accommodate possible explicit violation of $\mathcal{B}$ and/or $\mathcal{L}$ since one starts from the full symmetry content of the GUT-scale model.

The results of this paper should have implications in other settings. For instance, it is well known that the difference between the number density of baryons and that of antibaryons is about $10^{-10}$ when normalized to the entropy density of the Universe. To achieve this imbalance, some $\mathcal{B}$ and/or $\mathcal{L}$ violation appear compulsory. Whether it comes from non-trivial electroweak field configurations or from explicit $\mathcal{B}$ and/or $\mathcal{L}$ violating interactions, axions should be expected to participate since most of these mechanisms are not PQ-neutral. This is particularly true in GUT models, where the $\mathcal{B}+\mathcal{L}$ electroweak instanton interactions necessarily carry the same PQ charge as their strong counterparts used to solve the strong CP puzzle.

\begin{table}[p]
\begin{tabular}
[c]{llc}\hline
$\mathcal{B}-\mathcal{L}$ & Minimal $SU(5)$ models & $\beta$\\\hline
Exact & $\left\{
\begin{array}[c]{l}
\text{PQ~(\ref{PQFinalCharge})}\\
\text{Singlet DFSZ~(\ref{DFSZSU5})}\\
\text{Adjoint DFSZ~(\ref{PQaDFSZ})}%
\end{array}\right.$   
& Free\\
& $\rightarrow\ $SU(5)-invariant PQ charges~(\ref{PQSU5naif2}): & $\dfrac
{x}{2}-\dfrac{1}{x}\vspace{0.2007pc}$\\\hline
Broken & Type I seesaw $\left\{
\begin{array}[c]{l}
\text{PQ~(\ref{PQseesaw})}\\
\text{Singlet DFSZ~(\ref{TypeIU1W})}\\
\text{Adjoint DFSZ~(\ref{TypeIU1W})}%
\end{array}
\right.$ & $-x$, $\dfrac{1}{x}$\\
& Type II seesaw, PQ~(\ref{TypeIIU1W})
& $\dfrac{1}{2x}-\dfrac{x}{2}\vspace{0.2014pc}\vspace{0.2022pc}$\\
& Type II seesaw, 
$\left\{
\begin{array}[c]{l}
\text{Singlet DFSZ~(\ref{TypeIIU1W})}\\
\text{Adjoint DFSZ~(\ref{TypeIIU1W})}%
\end{array}
\right.$ &
$\dfrac{1}{2x}-\dfrac{x}{2}$, $\dfrac{1}{4x}-\dfrac{3x}{4}$, $\dfrac
{3}{4x}-\dfrac{x}{4}\vspace{0.2022pc}\vspace{0.203pc}$\\
& $\nu$DFSZ~(\ref{TablenuDFSZ}) & $\dfrac{1}{4x}-\dfrac{3x}{4}$, $\dfrac{5}%
{4x}+\dfrac{x}{4}$, $-\dfrac{1}{4x}-\dfrac{5x}{4}$, $\dfrac{3}{4x}%
-\dfrac{x}{4}\vspace{0.2022pc}\vspace{0.203pc}$\\\hline
$\mathcal{B}-\mathcal{L}$ & Flipped $SU(5)$ models & $\beta$\\\hline
Exact & $\left\{
\begin{array}[c]{l}%
\text{PQ}~(\ref{PQflippedSol})\\
\text{Singlet DFSZ~(\ref{FlipSinglet})}\\
\text{Fundamental DFSZ~(\ref{FlipFund})}%
\end{array}
\right.$ & Free\\
& $\rightarrow\ $SU(5)-invariant PQ charges~(\ref{PQflipSU5}): 
& $-x+\dfrac{1}{2x}\vspace{0.2014pc}%
$\\\hline
Broken & \multicolumn{1}{l}{PQ~(\ref{betagammaPQ})} & $-x\vspace{0.2022pc}\vspace{0.203pc}$\\
& Singlet DFSZ~(\ref{flipnDFSZ}) &  $\dfrac{1}{4x}-\dfrac{3x}{4}\vspace{0.2022pc}\vspace{0.203pc}$\\
& Fundamental DFSZ~(\ref{FundDFSZ}) & Many possibilities\\\hline
\end{tabular}
\caption{Summary of the values of the $\beta$ parameter, needed to compute the fermion PQ charges for various models analysed in the text.}
\label{Tablesum}
\end{table}

\newpage

\appendix

\section{Fermion masses}\label{AppMass}

The fermion masses are not correctly reproduced in the minimal $SU(5)$ model. The situation is similar in the PQ and DFSZ axion models built in that context since introducing a second Higgs fiveplets does not alter the relationship between $m_{d}$ and $m_{e}$, which both derive from the same Yukawa coupling. To cure for this, there are two solutions. The first is to introduce a new scalar multiplet. To induce non-trivial corrections to the Yukawa couplings, this multiplet cannot transform as $\mathbf{5}$, and by inspection, the simplest choice is as $\mathbf{45}$~\cite{Frampton:1979wf,Georgi:1979df}. The second solution is to introduce higher-dimensional operators, involving both $h_\mathbf{5}$ and $\mathbf{H}_\mathbf{24}$, thus exploiting the fact that $\mathbf{45} \subset \mathbf{5}\otimes\mathbf{24}$. The goal of this appendix is to present, briefly and in general terms, how these two solutions can be adapted to the PQ and DFSZ models.

\subsection{The $h_{\mathbf{45}}$ and axions}

Let us first consider the introduction of the $h_{\mathbf{45}}$. Concentrating on the fermion masses, two new Yukawa couplings can be constructed%
\begin{equation}
\mathcal{L}_{\text{Yukawa}}^{\mathbf{45}}=-\sqrt{3/8}\varepsilon_{ABCDE}%
(\bar{\chi}_{\mathbf{10}}^{c})^{AB}\mathbf{Y}_{10}^{\prime}%
(\chi_{\mathbf{10}})^{CF}(h_{\mathbf{45}})_{F}^{DE}-\sqrt{12}(\bar{\psi
}_{\mathbf{\bar{5}}}^{c})_{C}\mathbf{Y}_{5}^{\prime}(\chi
_{\mathbf{10}})^{AB}(h_{\mathbf{45}}^{\dagger})_{AB}^{C}+h.c.\;,
\label{Yuk45}%
\end{equation}
and they permit to induce reasonable fermion masses once $SU(5)$ breaks down. When plugging in the specific structure of the vacuum~\cite{Frampton:1979wf,Georgi:1979df,Kalyniak:1982pt,Eckert:1983bn}
\begin{equation}
(\mathbf{v}_{\mathbf{45}})_{C}^{AB}\equiv\langle0|(h_{\mathbf{45}})_{C}%
^{AB}|0\rangle=\frac{v_{45}}{4\sqrt{3}}(\delta_{C}^{A}-4\delta_{4}^{A}%
\delta_{C}^{4})\delta_{5}^{B}-(A\leftrightarrow B)\;,\label{vev45}%
\end{equation}
one finds in the one-fiveplet case,
\begin{subequations}
\label{C45yuk}%
\begin{align}
\mathbf{Y}_{u} &  =\mathbf{Y}_{10}\sin\alpha+\mathbf{Y}_{10}^{\prime}%
\cos\alpha,\;\;\mathbf{Y}_{10}=\mathbf{Y}_{10}^{T},\;\mathbf{Y}_{10}^{\prime
}=-\mathbf{Y}_{10}^{\prime T}\;,\\
\mathbf{Y}_{d} &  =\mathbf{Y}_{5}\sin\alpha+\mathbf{Y}_{5}^{\prime}\cos
\alpha,\;\mathbf{Y}_{e}=\mathbf{Y}_{5}^{T}\sin\alpha-3\mathbf{Y}_{5}^{\prime
T}\cos\alpha\;,
\end{align}
\end{subequations}
where $\tan\alpha=v_{5}/v_{45}$, $v^{2}=v_{5}^{2}+v_{45}^{2}\approx246\,$GeV, and $v\mathbf{Y}_{u,d,e}\rightarrow\sqrt{2}\mathbf{m}_{u,d,e}$ in the mass eigenstate basis.

For axion models, the $U(1)_{1}\otimes U(1)_{2}$ symmetry forces the $h_{\mathbf{45}}$ to be aligned with $h_{2,\mathbf{5}}$, with thus only the $\mathbf{Y}_{5}^{\prime}$ Yukawa coupling of Eq.~(\ref{Yuk45}) and no coupling to $\bar{\chi}_{\mathbf{10}}^{c}\chi_{\mathbf{10}}$. Also, because $h_{\mathbf{45}}$ is aligned with $h_{2,\mathbf{5}}$, the symmetry is not extended and there is no additional Goldstone boson. 

This may be a bit surprising because the potential can be constructed as invariant under $U(1)_{1}\otimes
U(1)_{2}\otimes U(1)_{45}$. What happens in that case is that when both $h_{\mathbf{45}}$ and $h_{2,\mathbf{5}}$ couple to the same fermion pair $\bar{\psi}_{\mathbf{\bar{5}}}^{c}\chi_{\mathbf{10}}$, we can construct the dimension-four effective potential term
\begin{equation}
\mathcal{L}_{eff}=c_{eff}\langle\mathbf{Y}_{10}^{\dagger}\mathbf{Y}%
_{10}\mathbf{Y}_{5}^{\dagger}\mathbf{Y}_{5}^{\prime}\rangle(h_{1,\mathbf{5}%
}^{\dagger})_{C}(h_{1,\mathbf{5}})^{B}(h_{2,\mathbf{5}})^{A}(h_{\mathbf{45}%
}^{\dagger})_{AB}^{C}+h.c.\ ,
\end{equation}
where we have used $(h_{\mathbf{45}})_{C}^{AB}=-(h_{\mathbf{45}})_{C}^{BA}$ and $(h_{\mathbf{45}})_{A}^{AB}=0$. Thus, the $U(1)_{1}\otimes U(1)_{2}\otimes U(1)_{45}$ symmetry cannot exist in the presence of the Yukawa couplings. Yet, this specific coupling cannot induce a mass for the pseudoscalar states because of the antisymmetry of $h_{\mathbf{45}}$. But with the help of $h_{1,\mathbf{5}}$ exchanges, the above coupling induces%
\begin{equation}
\mathcal{L}_{eff}\sim c_{eff}^{2}\langle\mathbf{Y}_{10}^{\dagger}%
\mathbf{Y}_{10}\mathbf{Y}_{5}^{\dagger}\mathbf{Y}_{5}^{\prime}\rangle
^{2}(h_{2,\mathbf{5}})^{A}(h_{\mathbf{45}}^{\dagger})_{AB}^{C}(h_{\mathbf{45}%
}^{\dagger})_{DC}^{B}(h_{2,\mathbf{5}})^{D}\ ,\label{PQEffPot}%
\end{equation}
which does contribute to pseudoscalar masses. To check that, let us plug in the polar representation of the scalar fields, with%
\begin{equation}
h_{\mathbf{45}}=\frac{1}{\sqrt{2}}\exp(i\eta_{45}/v_{45}%
)\mathbf{v}_{\mathbf{45}}+...\ ,\ h_{i,\mathbf{5}}=\frac{1}{\sqrt{2}}\exp
(i\eta_{i}/v_{i})\mathbf{v}_{i,\mathbf{5}}+...\ ,\label{h5Polar}%
\end{equation}
with $\mathbf{v}_{\mathbf{45}}$ given in Eq.~(\ref{vev45}) and $\mathbf{v}_{i,\mathbf{5}}=(0,0,0,0,v_{i})^{T}$. Under the assumption that the scalar potential is invariant under $U(1)_{1}\otimes U(1)_{2}\otimes U(1)_{45}$ at leading order, only the term in Eq.~(\ref{PQEffPot}) contributes and gives%
\begin{equation}
V(\eta_{1},\eta_{2},\eta_{45})\sim c_{eff}^{2}\langle\mathbf{Y}_{10}^{\dagger
}\mathbf{Y}_{10}\mathbf{Y}_{5}^{\dagger}\mathbf{Y}_{5}^{\prime}\rangle^{2}%
\cos\left(  2\frac{\eta_{2}}{v_{2}}-2\frac{\eta_{45}}{v_{45}}\right)  \ .
\end{equation}
The massive pseudoscalar is thus the combination $\pi^{0}\sim\eta_{2}/v_{2}-\eta_{45}/v_{45}$. For the massless states, first note that the electrically neutral state in $h_{\mathbf{45}}$ belongs to a colorless
$SU(2)_{L}$ doublet with hypercharge $1$, as do those in $h_{1,\mathbf{5}}$ and $h_{2,\mathbf{5}}$. So, the WBG of the $Z$ boson must be the combination $G^{0}\sim v_{1}\eta_{1}+v_{2}\eta_{2}+v_{45}\eta_{45}$. Knowing $\pi^{0}$ and $G^{0}$, the mixing matrix is immediate to construct%
\begin{equation}
\left(
\begin{array}
[c]{c}%
G^{0}\\
a^{0}\\
\pi^{0}%
\end{array}
\right)  =\left(
\begin{array}
[c]{ccc}%
\sin\beta & \cos\beta\sin\alpha & \cos\beta\cos\alpha\\
\cos\beta & -\sin\beta\sin\alpha & -\sin\beta\cos\alpha\\
0 & \cos\alpha & -\sin\alpha
\end{array}
\right)  \left(
\begin{array}
[c]{c}%
\eta_{1}\\
\eta_{2}\\
\eta_{45}%
\end{array}
\right)  \ ,\
\end{equation}
where we define
\begin{equation}
\tan\alpha=\frac{v_{2}}{v_{45}}\ ,\ \tan\beta=\frac{v_{1}}{\sqrt{v_{2}%
^{2}+v_{45}^{2}}}\ .
\end{equation}
From this, the PQ charge of the weak doublets inside $h_{1,\mathbf{5}}$, $h_{2,\mathbf{5}}$, and $h_{\mathbf{45}}$ are found to be $x$, $-1/x$, and $-1/x$, respectively. Thus, fermions have the same PQ charges as without the $h_{\mathbf{45}}$. One can also check that their couplings to the axion are not altered. This is simplest to see in the linear representation, in which the non-derivative couplings of the pseudoscalar states always enter in the combination $v_{i}+\operatorname{Re}h_{i}^{0}+i\eta_{i}$. In particular, all
dependencies on $\alpha$ cancel out of the axion couplings to fermions, leaving the same mass-dependent couplings as in the THDM,
\begin{equation}
\mathcal{L}_{a^{0}f\bar{f}}=\left(  x\bar{u}_{L}\mathbf{m}_{u}u_{R}-\frac
{1}{x}\bar{d}_{L}\mathbf{m}_{d}d_{R}-\frac{1}{x}\bar{e}_{L}\mathbf{m}_{e}%
e_{R}\right)  \frac{a^{0}}{v}\;,
\end{equation}
with $x=1/\tan\beta$ and $v^{2}=v_{1}^{2}+v_{2}^{2}+v_{45}^{2}$.

\subsection{Effective Yukawa couplings and axions}

Alternatively to the $h_{\mathbf{45}}$, higher-dimensional operators can be introduced~\cite{Barbieri:1979hc}. The same can be done in PQ and DFSZ axion models, provided these operators are constructed as invariant under the $U(1)_1\otimes U(1)_2$ symmetry. For instance, in the PQ model or in the singlet DFSZ model, given the charges in Eq.~(\ref{U1U2charges}), the leading corrections to the fermion masses can arise from
\begin{align}
\mathcal{L}_{\text{Yukawa}}^{\dim-5} &  =-\frac{\sqrt{2}}{\Lambda
}(h_{2,\mathbf{5}}^{\dagger})_{A}(\mathbf{H}_{\mathbf{24}})_{C}^{B}(\bar{\psi
}_{\mathbf{\bar{5}}}^{c})_{B}\mathbf{Y}_{5}^{\prime}(\chi
_{\mathbf{10}})^{AC}+\frac{2}{\Lambda}\varepsilon_{ABCDE}(\bar{\chi
}_{\mathbf{10}}^{c})^{AB}\mathbf{Y}_{10}^{\prime}(\chi_{\mathbf{10}%
})^{CF}(h_{1,\mathbf{5}})^{D}(\mathbf{H}_{\mathbf{24}})_{F}^{E}\nonumber\\
&  \;\;\;\;+\frac{1}{\Lambda}(h_{2,\mathbf{5}}^{\dagger})_{C}(\mathbf{H}%
_{\mathbf{24}})_{A}^{C}(\bar{\psi}_{\mathbf{\bar{5}}}^{c}%
)_{B}\mathbf{Y}_{5}^{\prime\prime}(\chi_{\mathbf{10}})^{AB}+\frac{1}{\Lambda
}\varepsilon_{ABCDE}(\bar{\chi}_{\mathbf{10}}^{c})^{AB}%
\mathbf{Y}_{10}^{\prime\prime}(\chi_{\mathbf{10}})^{CD}(h_{1,\mathbf{5}}%
)^{F}(\mathbf{H}_{\mathbf{24}})_{F}^{E}\;.
\end{align}
The $\mathbf{Y}_{5}^{\prime\prime}$ and $\mathbf{Y}_{10}^{\prime\prime}$ corrections can be absorbed into the leading Yukawa couplings since $(h_{1,\mathbf{5}})^{A}(\mathbf{H}_{\mathbf{24}})_{A}^{B}$ transforms as
$\mathbf{5} $. The $\mathbf{Y}_{5}^{\prime}$ and $\mathbf{Y}_{10}^{\prime}$ represent genuine corrections, and give%
\begin{equation}
\mathcal{L}_{\text{Yukawa}}^{\dim-5}\overset{\text{SSB}}{\rightarrow}%
-\frac{v_{5}v_{24}}{\sqrt{2}\Lambda}\cos\beta(\bar{d}_{R}^{i}\mathbf{Y}%
_{5}^{\prime}d_{L}^{i}-\frac{3}{2}\bar{e}_{R}\mathbf{Y}_{5}^{\prime T}%
e_{L})-\frac{v_{5}v_{24}}{\sqrt{2}\Lambda}\sin\beta\bar{u}_{R}^{i}%
(4\mathbf{Y}_{10}^{\prime T}-\mathbf{Y}_{10}^{\prime})u_{L}^{i}+h.c.\;.
\end{equation}
Note that compared to adding the $h_{\mathbf{45}}$, effective operators in general alter both the down and up-type Yukawa couplings.

The situation is a bit different in the adjoint DFSZ model, because the $\mathbf{H}_{\mathbf{24}}$ has a $U(1)_{1}\otimes U(1)_{2}$ charge. As a result, no operator can be constructed at the dimension-five level, and the leading corrections arise rather at the dimension-six level, with for example%
\begin{equation}
\mathcal{L}_{\text{Yukawa}}^{\dim-6}=\frac{\sqrt{2}}{\Lambda^{2}}\bar{\psi
}_{\mathbf{\bar{5}}}^{c}\mathbf{Y}_{5}^{a}\chi_{\mathbf{10}%
}\mathbf{H}_{\mathbf{24}}^{2}h_{1,\mathbf{5}}^{\dagger}+\frac{\sqrt{2}%
}{\Lambda^{2}}\bar{\psi}_{\mathbf{\bar{5}}}^{c}\mathbf{Y}_{5}^{b}%
\chi_{\mathbf{10}}\mathbf{H}_{\mathbf{24}}^{\dagger}\mathbf{H}_{\mathbf{24}%
}h_{2,\mathbf{5}}^{\dagger}+... \ ,
\end{equation}
where for both terms, the $SU(5)$ indices can be contracted in four different ways. Provided $v_{24}$ is not too small compared to $\Lambda$, these corrections are as effective as those of dimension-five to correct the fermion mass relations.

\end{document}